\documentclass[letterpaper,twocolumn,10pt]{article}
\usepackage{usenix-2020-09}

\usepackage{amsmath}
\usepackage{amssymb}
\usepackage{booktabs}
\usepackage{tabularx}
\usepackage{multirow}
\usepackage{graphicx}
\usepackage{subcaption}
\usepackage{xspace}
\usepackage{enumitem}
\usepackage{algorithm}
\usepackage{algorithmic}
\usepackage{pifont}
\usepackage{xcolor}
\usepackage{colortbl}
\usepackage[most]{tcolorbox}
\newcommand{\wayrow}{\cellcolor{gray!20}}

\newtcolorbox{promptbox}[1]{
	enhanced,
	breakable,
	colback=white,
	colframe=black!65,
	colbacktitle=black!65,
	coltitle=white,
	coltext=black,
	fonttitle=\bfseries,
	fontupper=\small,
	title={#1},
	boxrule=0.7pt,
	arc=1mm,
	left=2mm,
	right=2mm,
	top=1.5mm,
	bottom=1.5mm
}

\newcommand{\sysname}{\textsc{Anchor}\xspace}

\newcommand{\eg}{\textit{e.g.}\xspace}

\begin{document}
	
	\date{}
	
	\title{\Large \bf Know the Normal, Track the Attack: Context-Grounded and Stateful LLM Investigation over System Provenance}
	
	\author{
		{\rmfamily\upshape Lijie Zheng, Ji He, Ying Wang, Huang Zhang, Yulong Shen}\\
		Xidian University
	}
	
	\maketitle
	
	\begin{abstract}
		{Provenance-based intrusion detection systems (PIDSs) can localize suspicious activity in large audit streams, but their outputs remain difficult to turn into coherent accounts of how attacks unfold. Directly using LLMs to interpret local anomalous subgraphs does not solve this problem, since such analyses lack deployment-specific knowledge of normal behavior and cannot reliably carry validated attack state across evidence fragments. Consequently, routine system activities may be assigned unsupported attack semantics, while temporally dispersed evidence may be overlooked or attributed to incorrect attack stages.}

		{We present \sysname, an investigation-oriented provenance system that co-designs evidence curation and context-grounded LLM reasoning. \sysname calibrates anomaly judgments separately by relation type and links anomalous windows through rare relation-role patterns, forming evidence queues that preserve causal structure, temporal boundaries, and cross-window continuity. It then investigates process-centered evidence using two complementary forms of context, a Deployment Context constructed from environment-specific interaction and object baselines together with high-risk security knowledge, and a Case Context maintained by a confidence-gated Attack-Tracking Cache that keeps the investigation stateful across windows. By correlating current evidence with high-confidence prior findings, \sysname incrementally reconstructs attack narratives organized by kill-chain stages.}

		{We evaluate \sysname on six DARPA Transparent Computing E3/E5 datasets across three operating systems. Controlled evidence-level and end-to-end comparisons show that \sysname improves overall IoC recovery and attack-stage attribution over state-of-the-art provenance-based baselines, with the advantage persisting under a fixed LLM backbone. \sysname processes a full audit day at dollar-level API cost, providing a practical path from detection-oriented provenance output to context-grounded, longitudinal attack investigation.}
		
	\end{abstract}
	
	\section{Introduction}
	\label{sec:intro}
	
	Advanced persistent threats (APTs) unfold slowly across an entire system, blending malicious actions into vast streams of benign activity. To counter them, provenance-based intrusion detection systems (PIDSs) build directed graphs from kernel-level audit logs, in which nodes are system entities (processes, files, network flows) and edges are their timestamped interactions, and they flag deviations from learned normal behavior as potential intrusions~\cite{unicorn,threatrace,kairos,orthrus,flash,magic,provfusion,nodlink}. Extensive research has pushed reported detection accuracy to near-perfect levels.

	Yet near-perfect detection accuracy has not translated into adoption. An industrial survey identifies interpretation cost as a key bottleneck preventing the deployment of provenance-based endpoint detection and response (P-EDR) systems~\cite{dong2023we}. Existing systems typically output anomaly scores, flagged graph elements, or attack subgraphs that still require manual review. Although these outputs identify potential anomalies, they do not directly explain what happened, why the associated activities are suspicious, or how the attack developed. Security analysts must still identify critical evidence among abundant benign activity, interpret interactions among system entities, and organize scattered findings into a coherent attack narrative. A \emph{semantic gap} therefore remains between detection output and the interpretable attack narratives required by security analysts.
	
	With their strong language understanding and reasoning capabilities, LLMs offer a natural means of narrowing this gap. OCR-APT~\cite{ocrapt} is the first end-to-end system to feed detector-labeled provenance subgraphs to an LLM and generate attack-stage-oriented investigation reports. However, our analysis of its released reports reveals two missing forms of context (\S\ref{subsec:llmhalluc}). The first is \emph{deployment context}. Without normal-behavior baselines from the target environment, the model interprets current activities using generic pre-trained knowledge alone, and the released reports contain unsupported attack interpretations of routine deployment activities. The second is \emph{case context}. When each evidence fragment is analyzed independently, attack entities, critical interactions, and stage progression accumulated during the investigation cannot inform the current judgment. Although the released reports identify several ground-truth IoCs, many of them are assigned to incorrect attack stages (\S\ref{subsec:rq1}), indicating that local evidence provides limited support for reliable stage attribution.
	
	Supporting sustained LLM attack investigation requires solving two coupled challenges. First, raw audit streams are massive, attack activities are dispersed among abundant benign interactions, and anomaly signals vary across interaction types. The evidence interface must therefore control candidate volume while preserving attack-relevant interactions, local provenance structure, and temporal continuity. Second, the investigator must interpret this evidence using both deployment and case contexts. Deployment context should characterize environment-specific normal behavior and provide attack-oriented knowledge, while case context should accumulate attack entities, behavioral associations, and stage progression as new evidence arrives. Because carrying every historical judgment forward can propagate early uncertainty, case updates must be controlled. These requirements call for a coordinated design that curates current evidence before jointly reasoning over deployment context and a controlled case context.

	To this end, we develop \sysname, an investigation-oriented provenance system that acts as an analyst-inspired, stateful LLM investigator, comprising an evidence curation layer and a dual-context investigation layer. The evidence curation layer serves as the investigator's evidence interface. It identifies suspicious interactions from large provenance streams, links cross-window activities through relation-role rarity, and constructs evidence queues that preserve local provenance structure, explicit temporal boundaries, and cross-window continuity, thereby limiting the dilution of attack-relevant evidence by benign activity.
	
	The dual-context investigation layer follows the contextual reasoning workflow of security analysts. For each current evidence fragment, \sysname first compares it with the Deployment Context, which combines normal interaction baselines, object baselines, and high-risk security knowledge to characterize both deployment-relative deviation and potential attack semantics. \sysname then correlates the current evidence with the Case Context provided by a queue-scoped, confidence-gated Attack-Tracking Cache, which maintains previously admitted high-confidence evidence, the accumulated attack story, and stage assignments. Through this repeated observe-compare-correlate-update process, the investigator ultimately produces stage-structured attack narratives across windows.

	We evaluate \sysname on six datasets from the DARPA Transparent Computing E3~\cite{darpa_e3} and E5~\cite{darpa_e5} programs, which span three operating systems and are widely used in prior PIDS research~\cite{magic,flash,bilot2025sometimes,ocrapt,kairos,orthrus}. Controlled evidence-level and end-to-end comparisons show that \sysname improves overall IoC recovery and attack-stage attribution over state-of-the-art provenance-based baselines, with the advantage persisting under the same LLM backbone as OCR-APT. Ablation results demonstrate that deployment-specific normalcy baselines are the primary contributor to attack narrative quality, while the Attack-Tracking Cache provides further improvements. At this level of investigation quality, the end-to-end investigation cost remains at the dollar level.
	
	\paragraph{Contributions}
	\begin{itemize}[leftmargin=*,nosep]
		\item We identify and characterize two missing forms of context in LLM-based attack investigation. The lack of deployment-specific normalcy baselines leads to unsupported attack interpretations of routine activities, while the absence of accumulated case context leaves temporally dispersed evidence weakly linked and attributed to incorrect attack stages (\S\ref{subsec:llmhalluc}).
		\item We design \sysname, an investigation-oriented provenance system that jointly improves anomaly evidence organization and LLM-based investigation. Its relation-aware anomaly detector constructs cross-window evidence queues that capture attack progression, while its LLM-based investigator grounds judgments in a retrieval-augmented Deployment Context and keeps the investigation stateful through a Case Context maintained by a queue-scoped, confidence-gated Attack-Tracking Cache, connecting validated evidence across windows and supporting incremental attack reconstruction (\S\ref{sec:detector}, \S\ref{sec:llm}).
		\item We conduct a comprehensive evaluation on six DARPA TC E3/E5 datasets, demonstrating improved IoC recovery and attack-stage attribution over state-of-the-art provenance-based baselines, the effectiveness of deployment-specific normalcy grounding and attack tracking, and dollar-level end-to-end investigation costs (\S\ref{sec:eval}).
		\item We open-source our implementation and all labels to benefit the community and encourage further improvements upon our approach.
	\end{itemize}
	
	\section{Background and Motivation}
	\label{sec:background}
	
	This section develops the two challenges identified in the introduction through a concrete analysis of existing work. We first examine existing PIDSs from the perspective of output granularity and analyze why their outputs cannot be directly handed to an LLM to bridge the semantic gap (\S\ref{subsec:granularity}). We then take OCR-APT, the first LLM-supported attack investigation system, as a case study to examine two concrete problems that arise when an LLM generates attack reports directly from anomalous evidence flagged by the detector (\S\ref{subsec:llmhalluc}). Finally, we distill the design principles derived from these findings (\S\ref{subsec:principle}).
	
	\subsection{Why Not Just Feed a PIDS to an LLM?}
	\label{subsec:granularity}
	
	To bridge the semantic gap, a natural first attempt is to take an existing system's output and hand it directly to an LLM investigator. In what follows, we examine existing PIDSs from the perspective of output granularity, analyzing why none of them produces evidence that an LLM can reliably reason over to close the semantic gap.
	
	\noindent\textbf{Graph-level detection.}
	StreamSpot~\cite{streamspot}, Unicorn~\cite{unicorn}, and EdgeTorrent~\cite{king2023edgetorrent} classify an entire provenance graph (or a sequence of graph embeddings) as benign or malicious, producing one score per graph. This offers zero localization. An LLM told only that ``this graph is anomalous'' has no specific events to reason about and cannot construct any narrative at all.
	
	\noindent\textbf{Node-level detection.}
	ThreaTrace~\cite{threatrace}, Flash~\cite{flash}, MAGIC~\cite{magic}, ProvFusion~\cite{provfusion}, R-CAID~\cite{goyal2024r}, and VELOX~\cite{bilot2025sometimes} flag anomalous entities at the node level (or edge level). Their output is a set of discrete anomalous nodes lacking causal associations and temporal context among them. An LLM given only a scattered collection of anomalous node labels cannot reconstruct the causal progression of an attack.
	
	\begin{table}[t]
		\centering
		\caption{Properties of detection output as investigation evidence. Caus.: causal association among flagged elements. Temp.: explicit temporal boundaries. Cross-W.: continuity across time windows. Narr.: analyst-readable narrative output.}
		\label{tab:capability}
		\smallskip
		\footnotesize
		\setlength{\tabcolsep}{2pt}
	\begin{tabular}{|l|l|cccc|}
		\hline
		\textbf{System} & \textbf{Granularity} & \textbf{Caus.} & \textbf{Temp.} & \textbf{Cross-W.} & \textbf{Narr.} \\
		\hline
		Unicorn~\cite{unicorn} & Graph & \ding{55} & \ding{55} & \ding{55} & \ding{55} \\
		\hline
		ThreaTrace~\cite{threatrace} & Node & \ding{55} & \ding{55} & \ding{55} & \ding{55} \\
		\hline
		MAGIC~\cite{magic} & Node & \ding{55} & \ding{55} & \ding{55} & \ding{55} \\
		\hline
		Flash~\cite{flash} & Node & \ding{55} & \ding{55} & \ding{55} & \ding{55} \\
		\hline
		VELOX~\cite{bilot2025sometimes} & Node & \ding{55} & \ding{55} & \ding{55} & \ding{55} \\
		\hline
		NodLink~\cite{nodlink} & Subgraph & \ding{51} & \ding{55} & \ding{55} & \ding{55} \\
		\hline
		Kairos~\cite{kairos} & Subgraph & \ding{51} & \ding{55} & \ding{51} & \ding{55} \\
		\hline
		Orthrus~\cite{orthrus} & Subgraph & \ding{51} & \ding{51} & \ding{55} & \ding{55} \\
		\hline
		OCR-APT~\cite{ocrapt} & Subgraph & \ding{51} & \ding{55} & \ding{55} & \ding{51} \\
		\hline
		\wayrow \sysname & \wayrow Queue & \wayrow \ding{51} & \wayrow \ding{51} & \wayrow \ding{51} & \wayrow \ding{51} \\
		\hline
	\end{tabular}
	\end{table}

	\noindent\textbf{Subgraph-level investigation.}
	Subgraph-level investigation is the most promising direction, in which systems extract causally related substructures from alert points to provide attack context for analysts. ATLAS~\cite{atlas} and NodLink~\cite{nodlink} recover attack-related subgraphs, Kairos~\cite{kairos} runs Louvain community detection over edges with high anomaly scores to produce summary graphs, and Orthrus~\cite{orthrus} reconstructs attack paths through dependency analysis, compressing the number of nodes requiring manual inspection from thousands to between 41 and 123 and achieving high-quality attribution. However, analysts must still manually review these subgraphs and translate them into understandable attack narratives, leaving the semantic gap unbridged. Handing such subgraphs directly to an LLM for this translation does not solve the problem either. Although causally connected, they are presented as static structures, in which the temporal order and stage progression of interactions must be inferred from tens or even hundreds of edges, and benign and malicious interactions are mixed together, diluting the critical attack interactions.
	
	In summary, evidence that can support the LLM in reconstructing attack progression should preserve causal associations, carry explicit temporal boundaries, and capture the continuous evolution of the attack across time windows, and no existing output form satisfies all these conditions at once. Table~\ref{tab:capability} summarizes where existing systems stand against these requirements, together with whether they emit an analyst-readable narrative.
	
	\subsection{The Necessity of Context-Grounded Investigation}
	\label{subsec:llmhalluc}
	
	As a further step, OCR-APT~\cite{ocrapt} introduces an LLM on top of subgraph-level investigation, translating anomalous subgraphs into human-readable attack narratives and taking an important step toward bridging the semantic gap. Specifically, OCR-APT first constructs causal subgraphs around anomalous nodes via graph database queries, then feeds these subgraphs to an LLM that generates investigation reports organized by APT kill-chain stages. However, in this pipeline the LLM judges each piece of evidence in isolation with only generic pre-trained knowledge, lacking both a baseline for what constitutes normal behavior in the target deployment and access to the previously confirmed attack state.
	
	We analyze the official public reports released with OCR-APT on the DARPA TC E3 datasets and identify two characteristic problems.
	
	\noindent\textbf{Unsupported attack interpretations.} Lacking a normalcy baseline for the deployment environment, the LLM misclassifies standard FreeBSD monitoring activities (e.g., \texttt{vmstat}, \texttt{lsof}) and routine system operations (e.g., \texttt{devctl} reads and \texttt{pkg} auditing) as attack reconnaissance, fabricating coherent attack narratives around these benign activities. Table~\ref{tab:deployment_context_motivation} summarizes representative examples from the released reports.
	
	\begin{table}[t]
		\centering
		\caption{Representative deployment-ungrounded interpretations in OCR-APT's released E3-CADETS reports.}
		\label{tab:deployment_context_motivation}
		\smallskip
		\scriptsize
		\setlength{\tabcolsep}{2pt}
		\renewcommand{\arraystretch}{1.12}
		\begin{tabularx}{0.98\columnwidth}{|p{0.25\columnwidth}|X|}
			\hline
			\textbf{Activity} & \textbf{Reported interpretation} \\
			\hline
			\texttt{vmstat} & Internal reconnaissance and system-information collection \\
			\hline
			\texttt{lsof} & Reconnaissance, collection, and possible exfiltration \\
			\hline
			\texttt{devctl} & System reconnaissance and command-and-control activity \\
			\hline
			\texttt{410.pkg-audit} & Reconnaissance and possible trace removal \\
			\hline
			\texttt{411.pkg-backup} & Fileless execution, privilege escalation, and persistence \\
			\hline
		\end{tabularx}
	\end{table}
	
	\noindent\textbf{Limited IoC recovery and stage attribution.} Taking the publicly released IoC labels as ground truth, OCR-APT's final report correctly attributes only 3 out of 16 known IoCs on E3-CADETS to their corresponding attack stages (3/16), and 3 out of 7 on E3-THEIA (3/7). Genuine IoCs are buried among a large volume of benign activities flagged as anomalous and are thus hard to identify. Moreover, each judgment relies only on local subgraph context, without access to previously confirmed attack entities and stage assignments, leaving limited evidence for reliable stage attribution.
	
	These observations reveal two complementary limitations. Without deployment-specific baselines, the LLM may assign unsupported attack semantics to routine activities. Without accumulated case context, each judgment relies only on local evidence, providing limited support for reliable IoC identification and attack-stage attribution.
	
	\subsection{Design Principles}
	\label{subsec:principle}

Drawing on the above findings, we design \sysname around investigation-oriented evidence organization and context-grounded investigation.

\noindent\textbf{Investigation-oriented evidence organization.} We design a relation-aware anomaly detector that models different types of system interactions separately, reducing benign evidence introduced by heterogeneous anomaly scales. The detector organizes related anomalous activities into continuous cross-window queues, which are further divided into evidence fragments with explicit causal context and temporal boundaries, providing the LLM with a controlled set of candidate evidence for investigation.

\noindent\textbf{Deployment-context-grounded interpretation.} We construct normal interaction and object-behavior baselines from benign operation in the target environment and combine them with high-risk security knowledge to form the Deployment Context. Through retrieval augmentation, relevant context is supplied for the current evidence, allowing the LLM to jointly assess deployment-relative abnormality and potential attack semantics while reducing unsupported attack interpretations.

\noindent\textbf{Case Context through the Attack-Tracking Cache.} We design an Attack-Tracking Cache that maintains isolated, queue-scoped case states for different candidate investigations. The cache records high-confidence attack evidence, critical interactions, the evolving attack story, and stage assignments. Only investigation results confirmed as highly suspicious can update the cache and affect subsequent judgments, limiting uncertainty propagation and improving the reliability of evidence association and stage attribution.

	\section{Threat Model}
	\label{sec:threat}
	
	Consistent with prior work~\cite{kairos,orthrus,flash,magic,provfusion,ocrapt}, we consider APT adversaries whose objective is to compromise a target host and maintain unauthorized access or control over an extended period. Our analysis focuses on activities captured by standard kernel-level auditing frameworks~\cite{etw,linuxaudit,pasquier2017practical}, excluding threats that operate outside this scope, such as hardware-level side channels. We assume that the benign data used to train the detector and construct the normalcy baselines is collected in a trusted environment, thereby excluding data and model poisoning from our threat model. The trusted computing base includes the \sysname software, the provenance capture mechanism, and the underlying operating system, which are assumed to be protected from direct compromise by existing system-hardening techniques that are orthogonal and complementary to our work. Finally, we assume that the LLM service is not subject to adversarial manipulation~\cite{liu2023prompt,shi2023badgpt,zou2023universal,hubinger2024sleeper,zou2025poisonedrag}.
	
	\begin{figure*}[t]
		\centering
		\includegraphics[width=\textwidth]{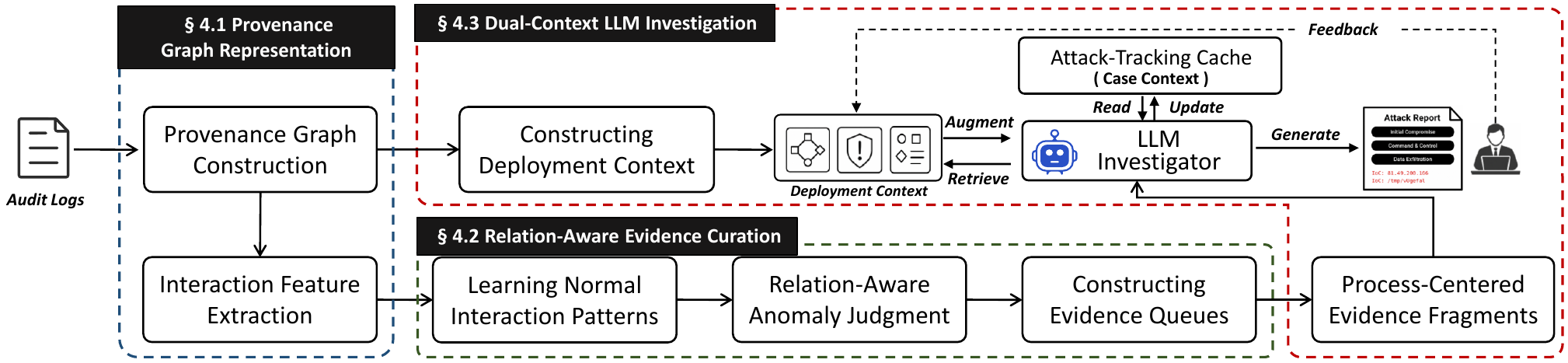}
		\caption{End-to-end architecture of \sysname.}
		\label{fig:overview}
	\end{figure*}

	\section{\sysname's Architecture}
	\label{sec:architecture}\label{sec:overview}
	
	Following the design principles in \S\ref{subsec:principle}, \sysname comprises an investigation-oriented evidence curation layer and a dual-context investigation layer. The evidence curation layer converts raw audit logs into provenance graphs, uses relation-aware anomaly detection to prioritize candidate evidence, and organizes related activities into continuous cross-window evidence queues, which are further divided into evidence fragments with explicit causal context and temporal boundaries. The dual-context investigation layer follows an analyst-inspired, stateful workflow. It observes the current evidence, compares it with normal-interaction baselines, object-behavior baselines, and high-risk security knowledge in the Deployment Context, and then correlates it with the Case Context maintained by the Attack-Tracking Cache to assess its attack semantics and stage roles. Only high-confidence investigation results can update the Case Context. By repeatedly observing, comparing, correlating, and updating, \sysname ultimately produces stage-structured attack narratives. Figure~\ref{fig:overview} shows the end-to-end architecture.

	\subsection{Provenance Graph Representation}
	\label{subsec:graph}
	
	This stage represents raw audit logs as a provenance graph and extracts interaction features for subsequent evidence curation and investigation.
	
	\subsubsection{Graph Construction}
	
	\sysname ingests audit logs from standard auditing frameworks (\eg ETW~\cite{etw}, Linux Audit~\cite{linuxaudit}, and CamFlow~\cite{pasquier2017practical}) and constructs a provenance graph $G=(V,E)$, where nodes represent system entities and edges represent timestamped causal interactions. Following prior work~\cite{nodlink,orthrus,bilot2025sometimes}, we model processes, files, and network flows with process-centered relations, as summarized in Table~\ref{tab:graph} (Appendix~\ref{sec:appendix_schema}). The event stream is partitioned into 15-minute windows~\cite{kairos} for subsequent anomaly detection.

\subsubsection{Interaction Feature Extraction}
	\label{subsec:features}
	
	The interactions between system entities are the natural unit for modeling deployment-specific normalcy. A process reading a configuration file and a process sending data to an external address represent fundamentally different behavioral patterns, each with its own notion of what is normal in a given deployment. To support this downstream modeling, we extract the features of each interaction relation into a feature vector composed of entity type and entity semantics. Specifically, for each interaction relation $e=(u,v,r_e,t_e)$, where $r_e$ denotes the relation type and $t_e$ the timestamp, we represent its features as $\mathbf{f}_e=[\mathbf{t}_u\|\mathbf{h}_u\|\mathbf{t}_v\|\mathbf{h}_v]$. $\mathbf{t}_u$ and $\mathbf{t}_v$ are one-hot vectors encoding the entity types of the source and destination nodes. $\mathbf{h}_u$ and $\mathbf{h}_v$ are dense semantic embeddings of the entities' textual attributes produced by Word2Vec~\cite{mikolov2013word2vec}, following prior work~\cite{orthrus,flash,provfusion}. The relation type $r_e$ is retained separately as a categorical edge type for relation-aware message passing and relation reconstruction.

	\subsection{Relation-Aware Evidence Curation}
	\label{subsec:model}\label{subsec:anomaly_section}\label{sec:detector}
	
\begin{figure*}[t]
		\centering
		\includegraphics[width=\textwidth]{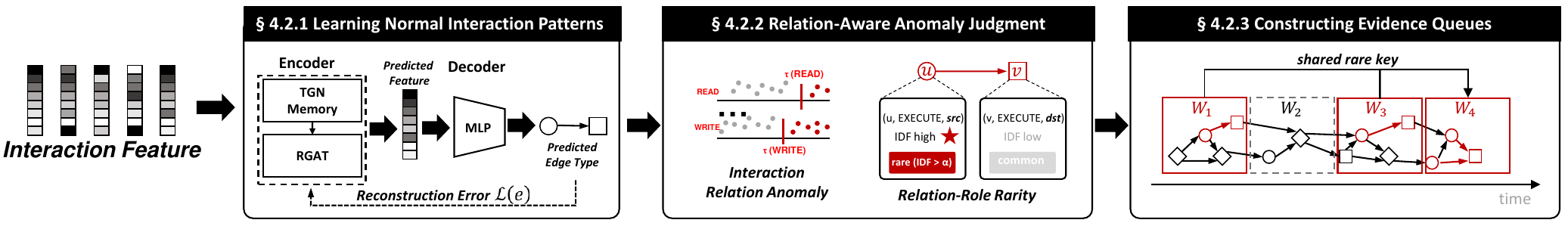}
		\caption{Overview of relation-aware evidence curation.}
		\label{fig:evidence_curation}
	\end{figure*}

	This stage filters, organizes, and connects candidate evidence for the investigator. Anomaly detection serves as the prioritization mechanism for candidate evidence. We first employ a dynamic graph learning approach over the extracted features to model normal interactions. We then perform relation-aware anomaly judgment in time-window units by comparing each interaction against the learned normal range of its relation type. Finally, we link windows sharing rare relation-role patterns into continuous evidence queues. {Figure~\ref{fig:evidence_curation} illustrates this pipeline.}
	
	\subsubsection{Learning Normal Interaction Patterns}
	
	To learn normal interaction patterns in the target deployment, \sysname trains a temporal relational encoder on benign audit data. A TGN memory module~\cite{tgn} maintains entity states that evolve with incoming interactions, while an RGAT encoder~\cite{rgatconv} aggregates neighborhood information conditioned on relation types. An MLP decoder reconstructs each interaction's relation type from its endpoint representations, and the model is trained by minimizing the corresponding cross-entropy loss $\mathcal{L}(e)$. At inference time, $\mathcal{L}(e)$ serves as an edge-level anomaly signal, with a higher loss indicating a larger deviation from the temporal and relational patterns learned from benign operation.
	
	\label{subsec:anomaly}
	
	\subsubsection{Relation-Aware Anomaly Judgment}
	
	This stage leverages the learned normal interaction patterns to detect anomalies. Whether an interaction is anomalous depends on how it deviates from the typical loss distribution of its relation type within the target deployment. We observe that different relation types exhibit distinct loss distributions under normal operation. For example, high-frequency \texttt{read} interactions are well learned by the model and produce consistently low losses, while infrequent \texttt{execute} interactions may yield higher losses even during normal operation. A uniform threshold across all relation types could miss genuine anomalies in low-loss types and produce false alarms in high-loss ones. We therefore ground anomaly judgment in the loss distribution of each relation type. Specifically, we identify anomalous interactions by combining per-relation anomaly scores with relation-role rarity, thereby reducing benign interaction noise.
	
	\noindent\textbf{Interaction-Relation Anomaly Judgment.}
	For each time window $W$ and relation type $r$, \sysname derives a relation-specific threshold $\tau(W,r)=\mu(W,r)+k\cdot\sigma(W,r)$, where $\mu(W,r)$ and $\sigma(W,r)$ denote the mean and standard deviation of the corresponding reconstruction losses, and $k$ is calibrated on benign data. An edge $e$ is marked as \emph{suspicious} when $\mathcal{L}(e)>\tau(W,r_e)$, and the window anomaly score is the average reconstruction loss of its suspicious edges.
	
	\noindent\textbf{Relation-Role Rarity Judgment.}
	To further characterize normal interactions in the target deployment, \sysname measures how rarely an entity appears under a specific relation and structural role. For example, a process may commonly act as the source of \texttt{read} events but rarely as the source of \texttt{execute} events. Following rarity-driven evidence prioritization~\cite{nodoze,priotracker,kairos}, for each suspicious edge $e=(u,v,r_e,t_e)$, we construct two relation-role keys, $(u,r_e,\texttt{src})$ and $(v,r_e,\texttt{dst})$, and compute $\mathrm{IDF}(key)=\log\frac{N}{\max(\mathrm{df}(key),1)}$ over $N$ benign calibration windows, where $\mathrm{df}(key)$ is the number of benign windows containing the key.
	
	\subsubsection{Constructing Evidence Queues}
	\label{subsec:queue}
	
	Based on suspicious edges and their relation-role rarity, \sysname constructs continuous evidence queues. A relation-role key participates in queue linkage only when its source edge exceeds the corresponding relation-specific threshold and its IDF exceeds a rarity threshold $\alpha$. Time windows sharing such keys are linked into the same \emph{evidence queue}. Since APT attacks are multi-stage campaigns whose transition periods may not produce high anomaly scores, we further incorporate intermediate windows when two windows in the same queue are separated by no more than a short gap $g$ (e.g., 30 minutes), providing a more complete evidence context for subsequent investigation. The queue score aggregates its window scores as $\text{score}(Q)=\sum_{W\in Q}\text{score}(W)$. A queue is submitted for subsequent investigation when its score exceeds an anomaly threshold $\beta$, calibrated on benign validation data following standard practice~\cite{kairos,sigl,orthrus}.

	\subsection{Dual-Context LLM Investigation}
	\label{subsec:online-investigation}\label{sec:llm}
	
	For each current evidence fragment extracted from the curated evidence queues, the LLM-based investigator follows a dual-context workflow. It first prepares and observes the current evidence, then compares it with the Deployment Context, correlates it with the Case Context maintained by the Attack-Tracking Cache, and uses confidence-gated admission to determine whether the investigation result may update the Case Context. This analyst-inspired process filters false positives and supports stage-structured investigation narratives while retaining cross-window continuity as an effect of maintained case state.

	\subsubsection{Constructing Deployment Context}
	
	The Deployment Context combines normal interaction baselines and object behavior baselines measured from benign operation with high-risk security knowledge derived from threat intelligence. The normal baselines characterize deployment-specific behavior, while the high-risk knowledge, maintained as a structured rule set, supplies security semantics for behaviors that require explicit scrutiny. Interaction and object baselines are embedded and stored in a vector database, where semantic retrieval serves as an implementation mechanism for selecting the context relevant to the current evidence.
	
	\noindent\textbf{Interaction Baselines.}
	Interaction baselines characterize how processes normally interact with other entities. We construct two categories, Process--Process and Process--NetFlow. For Process--NetFlow interactions, raw network addresses are highly sparse and carry little semantic information. Moreover, many attack behaviors, especially command-and-control communication, are more readily distinguished by whether communication leaves the local environment than by exact addresses. We therefore normalize network targets into internal and external address classes according to standard private address ranges. This representation reduces sparsity while preserving the internal-versus-external distinction critical for attack investigation.
	
	\noindent\textbf{High-Risk Knowledge.}
We do not build a benign Process--File interaction baseline. File interactions are substantially noisier than process or network interactions due to path variability, temporary artifacts, and application-specific identifiers. Instead, we maintain a separate high-risk Process--File rule set distilled from publicly available threat intelligence, aligned with MITRE ATT\&CK techniques~\cite{mitre_attack} such as credential access, persistence, and lateral movement. These rules are stored in structured form and supplied to the LLM as semantic attack context.
	
	\noindent\textbf{Object Baselines.}
	Object baselines provide compact normal profiles for normalized entity representations. Each object card records the entity type, normalized name, frequency level, common actions, common related objects, and a natural-language summary. While interaction baselines capture whether a behavioral pattern as a whole is normal, object baselines capture whether a specific process, file, or network object is common in the benign environment and whether its observed role is consistent with that environment.

	\subsubsection{Maintaining Case Context with the Attack-Tracking Cache}
	
	\begin{figure}[t]
		\centering
		\includegraphics[width=\columnwidth]{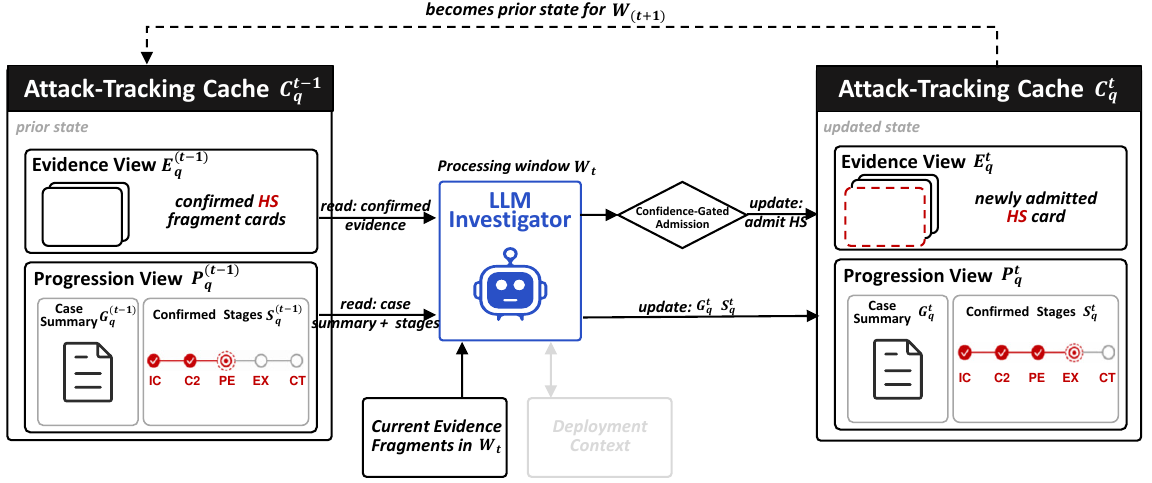}
		\caption{Interaction between the LLM investigator and the queue-scoped Attack-Tracking Cache when processing window $W_t$.}
		\label{fig:attack_tracking_cache}
	\end{figure}
	
	{For each evidence queue $q$, \sysname maintains its Case Context in an independent, queue-scoped Attack-Tracking Cache. We use $\mathcal{C}_{q}^{t}$ to denote the cache state after the queue has been processed through window $W_t$. The subscript $q$ identifies the evidence queue, and the superscript $t$ identifies the latest processed window. Figure~\ref{fig:attack_tracking_cache} illustrates how the LLM investigator reads and updates the cache when processing a window.}
	
	{The cache consists of an Evidence View and a Progression View, $\mathcal{C}_{q}^{t}=\big(\mathcal{E}_{q}^{t},\mathcal{P}_{q}^{t}\big)$. The Evidence View $\mathcal{E}_{q}^{t}$ stores the high-confidence attack fragment cards admitted through window $W_t$. Each card records its window identifier, seed process, fragment summary, suspicious and benign interactions, lifecycle-stage assignment, and supporting rationale. The Progression View is $\mathcal{P}_{q}^{t}=\big(\mathcal{G}_{q}^{t},\mathcal{S}_{q}^{t}\big)$, where $\mathcal{G}_{q}^{t}$ is the accumulated case summary and $\mathcal{S}_{q}^{t}$ is the confirmed attack-stage set. Together, the two views form the Case Context for the evidence queue.}
	
	{When processing $W_t$, \sysname reads previously confirmed attack evidence from $\mathcal{E}_{q}^{t-1}$ and uses it with the current evidence and Deployment Context in the LLM analysis. Only attack fragment cards finally labeled \texttt{highly\_suspicious} are admitted, so $\mathcal{E}_{q}^{t}=\mathcal{E}_{q}^{t-1}\cup\{f\in W_t\mid\operatorname{label}(f)=\mathrm{HS}\}$, where $\mathrm{HS}\equiv\texttt{highly\_suspicious}$. Results labeled \texttt{likely\_benign} (\texttt{LB}) or \texttt{suspicious} (\texttt{S}) are not written to the cache and therefore cannot influence subsequent investigation as historical attack evidence.}
	
	{\sysname aggregates the newly admitted attack fragment cards in $W_t$ into a window-level case summary, then combines it with $\mathcal{G}_{q}^{t-1}$ and $\mathcal{S}_{q}^{t-1}$ to update the Progression View. The updated Evidence View supplies confirmed historical attack evidence for subsequent fragment analysis, while the Progression View supplies the accumulated summary and stage information used for case-state updates and the stage-structured attack report. This recurrent read-and-update cycle makes the investigation stateful across windows by carrying confidence-gated evidence and accumulated attack progression from $\mathcal{C}_{q}^{t-1}$ to $\mathcal{C}_{q}^{t}$.}

	\subsubsection{Context-Grounded Attack Investigation}
	\label{subsec:online-pipeline}
	
\begin{figure}[t]
		\centering
		\includegraphics[width=\columnwidth]{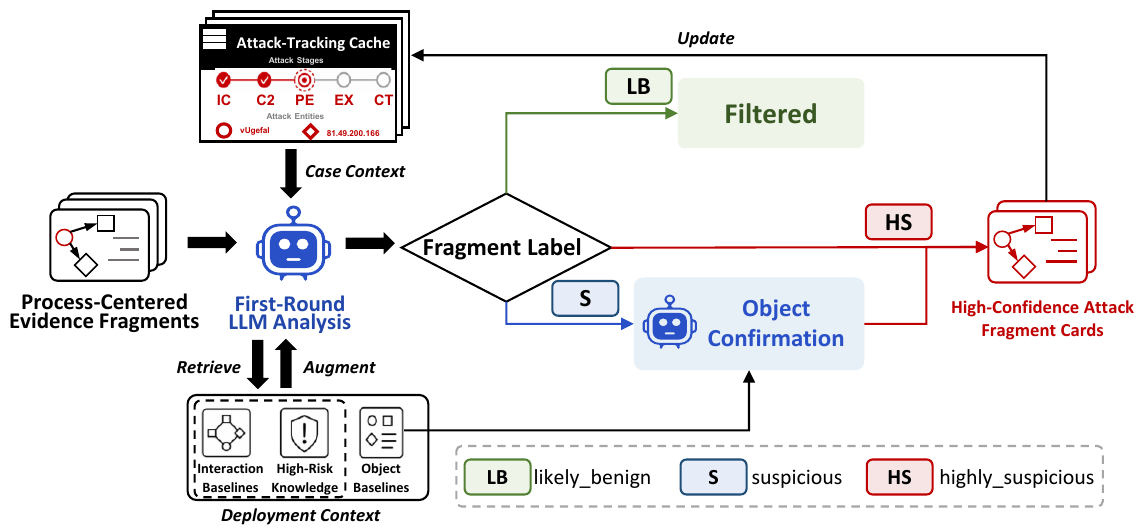}
		\caption{Context-grounded attack investigation with the Deployment Context and Case Context.}
		\label{fig:context_grounded_attack_investigation}
	\end{figure}

	For each candidate window prioritized by the evidence curation layer, \sysname converts local provenance activity into high-confidence attack fragment cards through current evidence preparation and two-round context-grounded analysis. Each resulting card records the attack evidence, lifecycle-stage assignment, and supporting rationale needed for subsequent case investigation. Figure~\ref{fig:context_grounded_attack_investigation} illustrates this process.
	
	\noindent\textbf{Current Evidence Preparation.}
	\sysname first reduces the provenance graph in each candidate window by merging edges with the same source, destination, and relation type while preserving their multiplicities and temporal ranges. Since security-relevant activity tends to form dense local regions centered on processes~\cite{nodoze,priotracker}, it then uses each process as a seed to extract a one-hop current evidence fragment. The fragment contains the seed process, time range, process, file, and network interactions and their multiplicities, and a template-generated summary for retrieving relevant context.
	
	\noindent\textbf{Two-Round Context-Grounded Analysis.}
	For each current evidence fragment, the first-round LLM analysis jointly uses the top-$k$ interaction baselines retrieved from the Deployment Context, the high-risk Process--File rules, and any existing Case Context available from the Attack-Tracking Cache. It produces a structured judgment containing the fragment label, suspicious and benign interactions, lifecycle-stage assignment, and supporting rationale. Fragments labeled \texttt{likely\_benign} are filtered, while those labeled \texttt{highly\_suspicious} are retained. A \texttt{suspicious} label triggers object confirmation, for which \sysname retrieves an object baseline for each object in the fragment. If any object lacks a normal baseline record, the fragment is upgraded directly to \texttt{highly\_suspicious}. Otherwise, the retrieved object baselines are added to the first-round context for a second LLM confirmation. Each fragment finally confirmed as \texttt{highly\_suspicious} is organized as a high-confidence attack fragment card and submitted to the Attack-Tracking Cache. The system prompt shared by both analysis rounds is provided as Core Prompt~1 in Appendix~\ref{sec:appendix_prompts}.
	
	\subsubsection{Stage-Structured Attack Report}
	\label{subsec:reconstruction}
	
	All fragments confirmed as \texttt{highly\_suspicious} are admitted to the Case Context. For each window, the LLM first synthesizes the admitted fragment cards into a concise window-level case summary capturing key processes, files, and network targets, including candidate IoCs, together with the attack stages covered. It then updates the case summary and confirmed stage set in the Attack-Tracking Cache and uses this Case Context to generate an investigation narrative organized by attack stages. The system prompt for this incremental update is provided as Core Prompt~2 in Appendix~\ref{sec:appendix_prompts}.
	
	The case narrative is organized according to the APT kill-chain model~\cite{holmes,ocrapt}, which decomposes an intrusion into eight narrative-oriented lifecycle stages spanning Initial Compromise (IC), Internal Reconnaissance (IR), Command and Control (C2), Privilege Escalation (PE), Lateral Movement (LM), Maintain Persistence (MP), Data Exfiltration (DE), and Covering Tracks (CT). The LLM assigns stages to individual fragments based on their semantic content and the accumulated Case Context, which are then aggregated into the stages covered by each window, without requiring an external classifier or rule engine. The confirmed stage assignments are maintained with the evolving case summary, providing stage-progression context for subsequent case updates. This maintained state allows the LLM to interpret later evidence in light of previously confirmed stages; for instance, once Initial Compromise and Command and Control are confirmed, the LLM can assess privilege-escalation evidence as a logical continuation of the established case narrative.
	
	Because only fragments judged highly suspicious through dual-context analysis enter this stage, the resulting narrative is built on confidence-gated evidence, helping reduce false-alarm contamination. As stage assignments accumulate in the Case Context, investigators gain continuous visibility into the kill-chain stages supported by the available evidence.

	\subsubsection{Feedback Mechanism}
	
	\sysname supports a feedback path from analyst decisions back into the Deployment Context. Fragment cards confirmed by analysts as true attacks can be incorporated into the high-risk rules repository, while false-alarm cases can be added to the normal interaction or object behavior baselines, strengthening the Deployment Context over time. This feedback mechanism provides a practical path for long-term adaptation as the deployment environment evolves.
	
	\section{Evaluation}
	\label{sec:eval}
	
	\noindent In this section, we systematically evaluate the investigation and evidence-curation capabilities of \sysname on six DARPA TC datasets through six research questions.
	
	\noindent\textbf{RQ1: Dual-Context Investigation Quality.} Does \sysname's Dual-Context Investigation, using both the Deployment Context and the Case Context maintained by the Attack-Tracking Cache, produce more complete and accurate attack narratives than the directly comparable end-to-end system?
	
	\noindent\textbf{RQ2: Impact of Investigation Evidence.} How do the quality and granularity of investigation evidence affect the final attack narrative?
	
	\noindent\textbf{RQ3: Evidence Curation Effectiveness.} Does \sysname's investigation-oriented evidence curation preserve attack windows and generate manageable evidence queues for candidate investigations?
	
	\noindent\textbf{RQ4: Investigation Cost.} What are the token usage and dollar cost of LLM-based investigation?
	
	\noindent\textbf{RQ5: Ablation.} How do the evidence-curation architecture, relation-aware anomaly judgment, Deployment Context, and Case Context components affect the results?
	
	\noindent\textbf{RQ6: Hyperparameter Sensitivity.} How sensitive is \sysname to key hyperparameters?
	
	\noindent\textbf{Implementation.}
	\sysname is implemented in Python. The detector uses a TGN~\cite{tgn} memory module with a two-layer RGAT~\cite{rgatconv} encoder and an MLP relation decoder, trained to reconstruct edge relation types on benign data. Node features are Word2Vec~\cite{mikolov2013word2vec} embeddings of entity text. The online investigator retrieves from a Qdrant~\cite{qdrant} vector store of interaction and object baselines and uses DeepSeek V4 Flash~\cite{xu2026deepseek} as its default backbone. We additionally evaluate GPT-4o-mini in RQ1 for a controlled backbone comparison. Unless stated otherwise, ablations and sensitivity studies are run on E3-CADETS. All experiments were performed on a server running Ubuntu 22.04, equipped with an Intel Xeon Platinum 8470Q processor, 256\,GB of RAM, and an NVIDIA RTX 5090 GPU.
	
	\subsection{Experimental Setup}
	\label{subsec:setup}
	
	\noindent\textbf{Datasets.}
	We leverage the publicly available datasets from the DARPA Transparent Computing (TC) program~\cite{darpa_e3,darpa_e5}, which are widely used as benchmarks for provenance-based intrusion detection~\cite{magic,flash,bilot2025sometimes,ocrapt,kairos,orthrus}. In the TC program, a team of security experts conducted adversarial engagements simulating realistic multi-stage attacks in an enterprise-like test environment, while scripted benign activity and background traffic continued throughout the engagements. We evaluate on six datasets spanning Engagement 3 (E3) and Engagement 5 (E5), collected by three TA1 provenance-capture systems in three distinct OS environments: CADETS (FreeBSD), THEIA (Linux), and ClearScope (Android). This setting lets us evaluate \sysname across heterogeneous deployment environments and engagement periods. For each dataset, we train the detector and construct deployment normalcy baselines from designated benign data, and evaluate on held-out days that include the report-documented attack activity. We give further details about how we split the datasets in Appendix~\ref{sec:appendix_split}.
	
	\noindent\textbf{Ground-Truth Labels.}
	For window-level labels in the detection task, we start from the attack-window labels provided in the public KAIROS implementation~\cite{kairos}. After cross-checking them against the official engagement reports, we apply necessary corrections and supplements to four datasets so that the evaluation windows better align with the attack timelines documented in the reports (Appendix~\ref{sec:appendix_labels}). For IoC labels in the investigation task, we use the public annotations from OCR-APT~\cite{ocrapt} on E3-CADETS and E3-THEIA, and for the remaining four datasets we carefully extract IoC labels item by item from the official engagement reports (the TA5.1 Ground Truth Report for E3~\cite{tc_gt_e3} and the TA5.1 Final Report for E5~\cite{ta51_final_e5}). Stage-attribution labels are likewise derived from these official reports. All of our labels are released with the open-source code.
	
	\noindent\textbf{Baselines.}
	For Dual-Context Investigation quality (RQ1), we perform an end-to-end comparison with OCR-APT~\cite{ocrapt}, the prior system most directly comparable to \sysname because it generates LLM-based attack narratives from detected provenance evidence, on the two datasets shared by its original evaluation and ours, namely E3-CADETS and E3-THEIA. To isolate model capacity from grounded evidence, we further re-run \sysname with GPT-4o-mini, the backbone used by OCR-APT, on these two datasets. For the impact of investigation evidence (RQ2), we convert the reproduced detection outputs of Orthrus~\cite{orthrus}, Kairos, MAGIC~\cite{magic}, Flash~\cite{flash}, and VELOX~\cite{bilot2025sometimes} into a common process-centered one-hop evidence format, feed them into the fixed Dual-Context Investigation protocol with the same dataset-specific Deployment Context used by \sysname, and evaluate the resulting narratives under a common protocol. This setup is justified because these systems do not themselves produce analyst-readable attack narratives, so measuring their investigation quality directly is inappropriate; holding the Dual-Context Investigation protocol fixed while varying only the upstream investigation evidence lets us analyze how the quality and granularity of investigation evidence affect the final narrative. To keep comparisons fair with each system's original evaluation, we report MAGIC and Flash on E3-CADETS and E3-THEIA, and Orthrus, Kairos, and VELOX on all six datasets. We exclude graph-level detectors such as StreamSpot~\cite{streamspot}, Unicorn~\cite{unicorn}, and EdgeTorrent~\cite{king2023edgetorrent}, which assign only a graph-wide anomaly decision without localizing attack events and thus provide no evidence an LLM can reason over (\S\ref{subsec:granularity}). We also exclude ProvFusion~\cite{provfusion} and SLOT~\cite{qiao2025slot}, because the partial ProvFusion release lacks the complete training and evaluation pipeline needed for reproduction, while an official implementation of SLOT was not publicly available at the time of our evaluation. For evidence curation effectiveness (RQ3), we compare against Kairos~\cite{kairos}, a window-level anomaly detector.

	\begin{table}[t]
		\centering
		\caption{End-to-end investigation comparison between \sysname and OCR-APT.}
		\label{tab:rq1_investigation}
		\smallskip
		\small
		\resizebox{\columnwidth}{!}{%
			\begin{tabular}{@{}|l|l|ccc|@{}}
				\hline
				\textbf{LLM} & \textbf{Dataset} & \textbf{System} & \textbf{IoC Cov. $\uparrow$} & \textbf{Stage Acc. $\uparrow$} \\
				\hline
				\multirow{4}{*}{DeepSeek V4 Flash}
				& \multirow{2}{*}{E3-CADETS} & \wayrow \sysname & \wayrow \textbf{14/16 (0.875)} & \wayrow \textbf{6/16 (0.375)} \\
				& & OCR-APT & 7/16 (0.438) & 4/16 (0.250) \\
				\cline{2-5}
				& \multirow{2}{*}{E3-THEIA} & \wayrow \sysname & \wayrow \textbf{7/7 (1.000)} & \wayrow \textbf{4/7 (0.571)} \\
				& & OCR-APT & 5/7 (0.714) & \textbf{4/7 (0.571)} \\
				\hline
				\multirow{4}{*}{GPT-4o-mini}
				& \multirow{2}{*}{E3-CADETS} & \wayrow \sysname & \wayrow \textbf{14/16 (0.875)} & \wayrow \textbf{6/16 (0.375)} \\
				& & OCR-APT & 11/16 (0.688) & 3/16 (0.188) \\
				\cline{2-5}
				& \multirow{2}{*}{E3-THEIA} & \wayrow \sysname & \wayrow \textbf{6/7 (0.857)} & \wayrow \textbf{4/7 (0.571)} \\
				& & OCR-APT & 5/7 (0.714) & 3/7 (0.429) \\
				\hline
			\end{tabular}%
		}
	\end{table}

	\noindent\textbf{Metrics.}
	RQ1 and RQ2 evaluate attack narratives with two complementary metrics. \emph{IoC coverage} (\textbf{IoC Cov.}) follows the definition in OCR-APT~\cite{ocrapt} and measures the fraction of ground-truth indicators of compromise recovered by the investigation pipeline. IoC coverage alone is insufficient to assess report quality, because investigation also requires placing recovered evidence in the correct attack stages. Fairly comparing free-text investigation reports is difficult, so we define \emph{stage-attribution accuracy} (\textbf{Stage Acc.}) as the fraction of all ground-truth typed IoCs that are recovered and attributed to one of their ground-truth kill-chain stages. This metric captures the correctness of IoC-to-stage attribution, a key dimension of investigation-report quality. Correctly locating evidence on the kill chain helps analysts understand attack progression and informs subsequent defensive actions~\cite{hutchins2011intelligence,holmes}. RQ3 evaluates the detector as a \emph{candidate generator} at the time-window level. The definitions of true positives (TP), false negatives (FN), false positives (FP), and true negatives (TN) follow Kairos~\cite{kairos}. We report precision, recall, and F1, where $\mathrm{Precision}=\frac{TP}{TP+FP}$, $\mathrm{Recall}=\frac{TP}{TP+FN}$, and $\mathrm{F1}=\frac{2\,\mathrm{P}\cdot\mathrm{R}}{\mathrm{P}+\mathrm{R}}$. As a candidate generator, RQ3 treats recall as the primary metric and precision as secondary, because missed attack windows reduce the evidence available to downstream investigation, while residual benign windows can still be filtered by the grounded investigator downstream. F1 summarizes the trade-off between recall and precision and supports overall comparisons among detectors in this candidate-generation role.

	\subsection{RQ1: Dual-Context Investigation Quality}
	\label{subsec:rq1}
	
	RQ1 assesses the end-to-end attack narrative quality of Dual-Context Investigation, which uses both the Deployment Context and the Case Context maintained by the Attack-Tracking Cache, compared with the directly comparable prior system. We compare against OCR-APT~\cite{ocrapt}, the prior system most directly comparable to \sysname because it generates LLM-based attack narratives from detected provenance evidence, on the two datasets shared by its original evaluation and ours, namely E3-CADETS and E3-THEIA. Narrative quality is measured by IoC coverage and stage-attribution accuracy (\S\ref{subsec:setup}). For OCR-APT, the DeepSeek rows are obtained from our rerun of its released pipeline, whereas the GPT-4o-mini rows are evaluated from its officially released recovered reports. Table~\ref{tab:rq1_investigation} reports both metrics under our default backbone, DeepSeek V4 Flash, and under the GPT-4o-mini backbone used by OCR-APT.
	
	In our DeepSeek V4 Flash reruns, \sysname improves IoC coverage from 0.438 to 0.875 on E3-CADETS and raises stage-attribution accuracy from 0.250 to 0.375. On E3-THEIA, IoC coverage improves from 0.714 to 1.000, while stage-attribution accuracy matches OCR-APT at 0.571. Overall, on both shared datasets, \sysname recovers more ground-truth IoCs and achieves stage-attribution accuracy that matches or exceeds OCR-APT.
	
	To reduce the effect of model differences, we re-run \sysname with GPT-4o-mini, the backbone reported in the OCR-APT paper, on the same two datasets. Using the same reported LLM model, \sysname still attains higher IoC coverage, 0.875 vs.\ 0.688 on E3-CADETS and 0.857 vs.\ 0.714 on E3-THEIA, together with higher stage-attribution accuracy. This shows that, under a shared GPT-4o-mini backbone, \sysname recovers a larger fraction of ground-truth IoCs with more correct stage attribution. Therefore, RQ1 indicates that \sysname recovers more ground-truth IoCs on both shared datasets and matches or exceeds OCR-APT in stage attribution, with this measured advantage persisting when using the same reported LLM model. Appendix~\ref{sec:appendix_casestudy} further presents an E5-CADETS case study illustrating how \sysname organizes a two-day attack into a single kill-chain narrative.
	
	\begin{table}[t]
		\centering
		\caption{Impact of upstream investigation evidence under a fixed Dual-Context Investigation protocol.}
		\label{tab:rq2_detection_evidence}
		\smallskip
		\small
		\resizebox{\columnwidth}{!}{%
			\begin{tabular}{@{}|l|c|c|c|@{}}
				\hline
				\textbf{Dataset} & \textbf{System} & \textbf{IoC Cov. $\uparrow$} & \textbf{Stage Acc. $\uparrow$} \\
				\hline
				\multirow{6}{*}{E3-CADETS}
				& \wayrow \sysname & \wayrow \textbf{14/16 (0.875)} & \wayrow \textbf{6/16 (0.375)} \\
				& Orthrus & 12/16 (0.750) & 2/16 (0.125) \\
				& Kairos & 13/16 (0.813) & 3/16 (0.188) \\
				& MAGIC & 7/16 (0.438) & 0/16 (0.000) \\
				& Flash & 9/16 (0.563) & 5/16 (0.313) \\
				& VELOX & 10/16 (0.625) & 0/16 (0.000) \\
				\hline
				\multirow{4}{*}{E5-CADETS}
				& \wayrow \sysname & \wayrow 13/15 (0.867) & \wayrow \textbf{10/15 (0.667)} \\
				& Orthrus & 10/15 (0.667) & 1/15 (0.067) \\
				& Kairos & \textbf{14/15 (0.933)} & 5/15 (0.333) \\
				& VELOX & 9/15 (0.600) & 1/15 (0.067) \\
				\hline
				\multirow{6}{*}{E3-THEIA}
				& \wayrow \sysname & \wayrow \textbf{7/7 (1.000)} & \wayrow \textbf{4/7 (0.571)} \\
				& Orthrus & 5/7 (0.714) & 0/7 (0.000) \\
				& Kairos & 3/7 (0.429) & 0/7 (0.000) \\
				& MAGIC & 4/7 (0.571) & 3/7 (0.429) \\
				& Flash & 0/7 (0.000) & 0/7 (0.000) \\
				& VELOX & 4/7 (0.571) & 1/7 (0.143) \\
				\hline
				\multirow{4}{*}{E5-THEIA}
				& \wayrow \sysname & \wayrow \textbf{10/14 (0.714)} & \wayrow \textbf{7/14 (0.500)} \\
				& Orthrus & 8/14 (0.571) & 1/14 (0.071) \\
				& Kairos & \textbf{10/14 (0.714)} & 3/14 (0.214) \\
				& VELOX & 7/14 (0.500) & 2/14 (0.143) \\
				\hline
				\multirow{4}{*}{E3-ClearScope}
				& \wayrow \sysname & \wayrow \textbf{3/14 (0.214)} & \wayrow \textbf{2/14 (0.143)} \\
				& Orthrus & 2/14 (0.143) & 0/14 (0.000) \\
				& Kairos & 2/14 (0.143) & 0/14 (0.000) \\
				& VELOX & 1/14 (0.071) & 0/14 (0.000) \\
				\hline
				\multirow{4}{*}{E5-ClearScope}
				& \wayrow \sysname & \wayrow \textbf{7/18 (0.389)} & \wayrow \textbf{2/18 (0.111)} \\
				& Orthrus & 1/18 (0.056) & 0/18 (0.000) \\
				& Kairos & 6/18 (0.333) & \textbf{2/18 (0.111)} \\
				& VELOX & 0/18 (0.000) & 0/18 (0.000) \\
				\hline
			\end{tabular}%
		}
	\end{table}

	\subsection{RQ2: Impact of Investigation Evidence}
	\label{subsec:rq2}
	
	RQ2 examines how the quality and granularity of upstream investigation evidence affect the final attack narrative under a fixed Dual-Context Investigation protocol. We use the reproduced detection outputs of Orthrus~\cite{orthrus}, Kairos~\cite{kairos}, MAGIC~\cite{magic}, Flash~\cite{flash}, and VELOX~\cite{bilot2025sometimes} to select investigation windows and seed processes, from which the fixed Dual-Context Investigation protocol constructs process-centered one-hop fragments using the same dataset-specific Deployment Context as \sysname. MAGIC, Flash, and VELOX provide node-level anomaly evidence; we use the reconstructed attack-summary graphs of Orthrus and Kairos as subgraph-level evidence; and \sysname provides temporally ordered evidence queues. We evaluate the resulting narratives using the same IoC coverage and stage-attribution accuracy metrics defined for RQ1. Table~\ref{tab:rq2_detection_evidence} reports the results.
	
	\noindent\textbf{Node-level Evidence.} Node-level evidence struggles to support a complete narrative. MAGIC and Flash produce discrete anomalous node candidates and can therefore supply more seeds for one-hop expansion, recovering more IoCs on some datasets, such as 0.563 for Flash on E3-CADETS and 0.571 for MAGIC on E3-THEIA. The broader candidate set also admits many benign system processes and still often omits mid- and late-stage attack tools. VELOX replaces the complex temporal encoder with a lightweight linear model, yet still recovers a considerable share of IoCs (0.625 on E3-CADETS), showing that node-level signals suffice to hit part of the attack entities, while its stage-attribution accuracy stays close to zero (at most 0.143). Overall, discrete anomalous nodes lack causal and temporal organization and cannot stably support cross-stage narratives.
	
	\noindent\textbf{Subgraph-level Evidence.} Subgraph-level evidence is clearly stronger than node-level evidence. Kairos builds temporal window queues from edge anomalies and generates attack summary graphs, and Orthrus performs forward and backward dependency tracing within time windows around anomalous nodes to reconstruct attack summary graphs, so both recover a large share of ground-truth IoCs on some datasets (0.933 for Kairos on E5-CADETS and 0.750 for Orthrus on E3-CADETS). However, such evidence is essentially a static graph structure, in which the temporal order and stage progression of interactions are not explicitly organized, and the LLM must infer the sequence of the attack from tens or even hundreds of edges, so stage attribution remains generally low. This is consistent with the analysis in \S\ref{subsec:granularity}, where subgraphs preserve causal associations but do not explicitly organize temporal and stage progression, supporting IoC recovery more consistently than accurate stage attribution.
	
	\noindent\textbf{Overall Results.} Across the six datasets, \sysname's curated investigation evidence achieves the highest IoC coverage on four datasets, ties for the best on E5-THEIA, and attains the highest or tied-highest stage-attribution accuracy on every dataset. Because all outputs are converted into the same process-centered one-hop format and evaluated under the same downstream investigation protocol, the observed differences primarily reflect differences in the upstream detection evidence supplied to the investigator. Coverage drops on ClearScope, where the Android environment has sparse benign semantics and report-defined IoCs are dominated by short-lived APK artifacts, leaving limited room for grounding; even so, \sysname still matches or exceeds every alternative.
	
	\begin{table}[t]
		\centering
		\caption{Detection results of \sysname. Datasets marked with * use corrected ground-truth labels (Appendix~\ref{sec:appendix_labels}).}
		\label{tab:window}
		\smallskip
		\resizebox{\columnwidth}{!}{%
			\begin{tabular}{|l|c|c|c|c|c|c|c|}
				\hline
				\textbf{Dataset} & \textbf{TP} & \textbf{TN} & \textbf{FP $\downarrow$} & \textbf{FN $\downarrow$} & \textbf{Rec. $\uparrow$} & \textbf{Prec. $\uparrow$} & \textbf{F1 $\uparrow$} \\
				\hline
				E3-CADETS      & 4  & 177 & 0  & 0 & 1.000 & 1.000 & 1.000 \\
				\hline
				E5-CADETS*     & 12 & 234 & 11 & 0 & 1.000 & 0.522 & 0.686 \\
				\hline
				E3-ClearScope  & 5  & 107 & 8  & 0 & 1.000 & 0.385 & 0.556 \\
				\hline
				E5-ClearScope* & 18 & 255 & 13 & 0 & 1.000 & 0.581 & 0.735 \\
				\hline
				E3-THEIA*      & 7  & 214 & 1  & 3 & 0.700 & 0.875 & 0.778 \\
				\hline
				E5-THEIA*      & 2  & 180 & 3  & 0 & 1.000 & 0.400 & 0.571 \\
				\hline
			\end{tabular}%
		}
	\end{table}

	\subsection{RQ3: Evidence Curation Effectiveness}
	\label{subsec:rq3}
	
	RQ3 evaluates the detector as the candidate-generation mechanism of the evidence curation layer, treating recall as the primary metric and precision and F1 as measures of candidate-set accuracy. Table~\ref{tab:window} reports the window-level detection results of \sysname, and Table~\ref{tab:window_compare} compares \sysname with Kairos.
	
	\noindent\textbf{Detection Performance.} \sysname achieves a recall of $1.000$ on five of the six datasets and $0.700$ on E3-THEIA, showing that the detector retains all labeled attack windows as candidates on most datasets. The detector also selects some non-attack-labeled windows, reducing window-level precision on several datasets. These false positives primarily increase the number of candidate windows for downstream investigation, where the grounded investigator assesses them against the deployment's normal-behavior baselines and only evidence judged highly suspicious enters the final attack narrative.
	
	\noindent\textbf{FN Analysis.} The three FNs on E3-THEIA are not included in selected evidence queues because their anomalous evidence lacks sufficient rare relation-role key overlap with those queues, in two patterns. The first window mainly covers the maintenance and loss of a Drakon connection, a period that leaves limited explicit behavioral evidence in the audit logs. The other two windows contain explicit re-exploitation and payload-deployment activities, but the attack entities transition from \texttt{clean} to \texttt{profile} and \texttt{/var/log/xdev}, and from the Firefox browser extension and \texttt{loaderDrakon} to \texttt{gtcache} and \texttt{profile}, respectively, leaving adjacent windows without identical relation-role keys. Sparse explicit evidence and cross-stage changes in attack processes and files can therefore affect anomaly judgments based on relation-role key rarity.
	
	\noindent\textbf{Comparison with Kairos.} Table~\ref{tab:window_compare} compares \sysname with Kairos. To ensure a fair comparison, we rerun Kairos using the corrected dataset labels. Except on the ClearScope datasets, \sysname achieves precision no lower than Kairos. It also matches or exceeds Kairos in recall across all six datasets and achieves a higher F1 on five datasets. Kairos achieves higher precision on both ClearScope datasets. In ClearScope, both attack and abundant benign activities occur within Firefox and can share similar audit-level entities and interactions. Kairos evaluates rareness at the node level, whereas \sysname evaluates relation-role keys and can therefore distinguish interactions involving the same common entity.
	
	\begin{table}[t]
		\centering
		\caption{Comparison of detection results with Kairos.}
		\label{tab:window_compare}
		\smallskip
		\resizebox{\columnwidth}{!}{%
			\begin{tabular}{|l|l|c|c|c|}
				\hline
				\textbf{Dataset} & \textbf{System} & \textbf{Recall $\uparrow$} & \textbf{Precision $\uparrow$} & \textbf{F1 $\uparrow$} \\
				\hline
				\multirow{2}{*}{E3-CADETS}
				& \wayrow \sysname & \wayrow \textbf{1.000} & \wayrow \textbf{1.000} & \wayrow \textbf{1.000} \\
				\cline{2-5}
				& Kairos  & \textbf{1.000} & 0.800 & 0.889 \\
				\hline
				\multirow{2}{*}{E5-CADETS*}
				& \wayrow \sysname & \wayrow \textbf{1.000} & \wayrow \textbf{0.522} & \wayrow \textbf{0.686} \\
				\cline{2-5}
				& Kairos  & 0.750 & 0.450 & 0.563 \\
				\hline
				\multirow{2}{*}{E3-ClearScope}
				& \wayrow \sysname & \wayrow \textbf{1.000} & \wayrow 0.385 & \wayrow 0.556 \\
				\cline{2-5}
				& Kairos  & \textbf{1.000} & \textbf{0.714} & \textbf{0.833} \\
				\hline
				\multirow{2}{*}{E5-ClearScope*}
				& \wayrow \sysname & \wayrow \textbf{1.000} & \wayrow 0.581 & \wayrow \textbf{0.735} \\
				\cline{2-5}
				& Kairos  & 0.722 & \textbf{0.722} & 0.722 \\
				\hline
				\multirow{2}{*}{E3-THEIA*}
				& \wayrow \sysname & \wayrow \textbf{0.700} & \wayrow \textbf{0.875} & \wayrow \textbf{0.778} \\
				\cline{2-5}
				& Kairos  & 0.300 & 0.600 & 0.400 \\
				\hline
				\multirow{2}{*}{E5-THEIA*}
				& \wayrow \sysname & \wayrow \textbf{1.000} & \wayrow \textbf{0.400} & \wayrow \textbf{0.571} \\
				\cline{2-5}
				& Kairos  & \textbf{1.000} & 0.031 & 0.060 \\
				\hline
			\end{tabular}%
		}
	\end{table}
	
	\begin{table}[t]
		\centering
		\caption{Investigation cost under DeepSeek V4 Flash pricing.}
		\label{tab:api_cost}
		\smallskip
		\small
		\resizebox{\columnwidth}{!}{%
			\begin{tabular}{|l|l|c|c|c|}
				\hline
				\textbf{Dataset} & \textbf{System} & \textbf{Input Tok.} & \textbf{Output Tok.} & \textbf{Cost (\$)} \\
				\hline
				\multirow{7}{*}{E3-CADETS}
				& \wayrow \sysname & \wayrow 4,464,212 & \wayrow 246,691 & \wayrow 0.6713 \\
				& Orthrus & 118,829 & 19,767 & 0.0212 \\
				& Kairos & 3,822,618 & 530,230 & 0.6527 \\
				& MAGIC & 3,286,142 & 231,834 & 0.5156 \\
				& Flash & 2,892,801 & 236,767 & 0.4314 \\
				& VELOX & 55,927 & 17,034 & 0.0071 \\
				& OCR-APT & 41,766 & 18,043 & 0.0067 \\
				\hline
				\multirow{7}{*}{E3-THEIA}
				& \wayrow \sysname & \wayrow 8,839,288 & \wayrow 430,564 & \wayrow 1.3100 \\
				& Orthrus & 779,591 & 60,437 & 0.1240 \\
				& Kairos & 5,903,225 & 300,383 & 0.8914 \\
				& MAGIC & 140,619 & 30,978 & 0.0191 \\
				& Flash & 6,850,707 & 448,699 & 1.0326 \\
				& VELOX & 80,233 & 28,495 & 0.0101 \\
				& OCR-APT & 57,859 & 24,302 & 0.0086 \\
				\hline
			\end{tabular}%
		}
	\end{table}
	
	\subsection{RQ4: Investigation Cost}
	\label{subsec:rq4}
	
	RQ4 evaluates the token usage and dollar cost of grounded investigation. We report input tokens, output tokens, and API cost under DeepSeek V4 Flash pricing on E3-CADETS and E3-THEIA. Table~\ref{tab:api_cost} includes the end-to-end investigation cost of OCR-APT as well as the investigation cost of feeding Orthrus, Kairos, MAGIC, Flash, and VELOX detection outputs into the same grounded investigator. \sysname completes a full-day investigation at \$0.67 on E3-CADETS and \$1.31 on E3-THEIA, remaining at the dollar level overall. Compared with the cent-level cost of OCR-APT, \sysname is more expensive because its normalcy retrieval, cache-conditioned fragment analysis, and incremental progression updates across windows require additional LLM processing. Read together with RQ1 and RQ2, this extra cost corresponds to higher narrative quality: on E3-CADETS, \sysname raises IoC coverage from OCR-APT's 0.438 to 0.875 and achieves stage-attribution quality that matches or exceeds the compared systems. Therefore, grounded investigation obtains a more complete attack narrative at an acceptable absolute cost, and end-to-end deployment remains economically practical.

	\subsection{RQ5: Ablation}
	\label{subsec:rq5}
	
	RQ5 uses replacement and incremental ablations to assess how the evidence-curation architecture, relation-aware anomaly judgment, Deployment Context, and Case Context affect the results on E3-CADETS.
	
	\noindent\textbf{Detector.} Table~\ref{tab:gnn_ablation} compares RGAT with RGCN~\cite{rgcn,ocrapt}, UniMP~\cite{unimp,kairos}, Orthrus-style~\cite{orthrus}, and GAT~\cite{gat,magic}. All variants keep recall at 1.000, so the main difference among candidate generators lies in false-positive control rather than missed attacks. RGAT and UniMP produce no false-positive windows, while RGCN, Orthrus-style, and GAT yield 11, 9, and 8, respectively. Since RGCN, UniMP, and Orthrus-style incorporate relation information differently, access to relation type alone does not determine false-positive control. RGAT further records 0.60\,s per window, approximately one quarter of the inference times of the alternatives, providing the best observed combination of candidate-set cleanliness and latency while preserving complete recall.
	
	\begin{table}[t]
		\centering
		\caption{Detector ablation on E3-CADETS. Time is the inference time averaged per time window over the 181 evaluation windows.}
		\label{tab:gnn_ablation}
		\smallskip
		\small
		\resizebox{\columnwidth}{!}{%
			\begin{tabular}{|l|c|c|c|c|c|c|}
				\hline
				\textbf{Encoder} & \textbf{TP} & \textbf{FP $\downarrow$} & \textbf{Rec. $\uparrow$} & \textbf{Prec. $\uparrow$} & \textbf{Time (s/win.) $\downarrow$} & \textbf{Mem. (MB) $\downarrow$} \\
				\hline
				\wayrow \textbf{RGAT} & \wayrow 4 & \wayrow \textbf{0} & \wayrow \textbf{1.000} & \wayrow \textbf{1.000} & \wayrow \textbf{0.60} & \wayrow 6174.19 \\
				\hline
				RGCN & 4 & 11 & \textbf{1.000} & 0.267 & 2.49 & 5833.99 \\
				\hline
				UniMP & 4 & \textbf{0} & \textbf{1.000} & \textbf{1.000} & 2.42 & 5833.74 \\
				\hline
				Orthrus-style & 4 & 9 & \textbf{1.000} & 0.308 & 2.42 & 5833.03 \\
				\hline
				GAT & 4 & 8 & \textbf{1.000} & 0.333 & 2.49 & \textbf{5831.17} \\
				\hline
			\end{tabular}%
		}
	\end{table}
	
\begin{figure}[t]
		\centering
		\includegraphics[width=\linewidth]{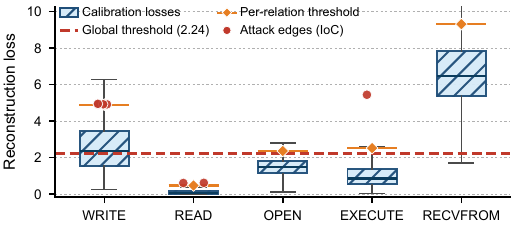}
		\caption{Per-relation reconstruction-loss distributions on E3-CADETS. Boxplots and thresholds are from the validation day 2018-04-05, and red points mark IoC-matched attack edges on 2018-04-06. Orange marks the per-relation threshold and the red dashed line marks the global threshold.}
		\label{fig:relation_loss_thresholds}
	\end{figure}

	\noindent\textbf{Anomaly Judgment.} Table~\ref{tab:relation_ablation} compares two anomaly-judgment strategies. Replacing the per-relation threshold and relation-role IDF with a global threshold and node-level IDF~\cite{kairos} keeps recall at 1.000, but precision drops from 1.000 to 0.031 and false positives rise from 0 to 124. Figure~\ref{fig:relation_loss_thresholds} explains why. Normal reconstruction losses differ substantially across relation types, for example, the median is about 0.08 for \texttt{READ} but about 6.48 for \texttt{RECVFROM}. A single global threshold therefore misses genuine anomalies in low-loss relations and flags normal behavior in high-loss relations as anomalous, and these false positives are exactly the benign interaction noise that dilutes the evidence queues.
	
	\begin{table}[t]
		\centering
		\caption{Anomaly-judgment ablation on E3-CADETS.}
		\label{tab:relation_ablation}
		\smallskip
		\small
		\begin{tabular}{|l|c|c|c|c|c|}
			\hline
			\textbf{Judgment} & \textbf{TP} & \textbf{FP $\downarrow$} & \textbf{TN} & \textbf{Rec. $\uparrow$} & \textbf{Prec. $\uparrow$} \\
			\hline
			Per-relation (ours) & 4 & \textbf{0} & 177 & 1.000 & \textbf{1.000} \\
			\hline
			Global threshold & 4 & 124 & 53 & 1.000 & 0.031 \\
			\hline
		\end{tabular}
	\end{table}
	
\begin{figure}[t]
		\centering
		\includegraphics[width=\linewidth]{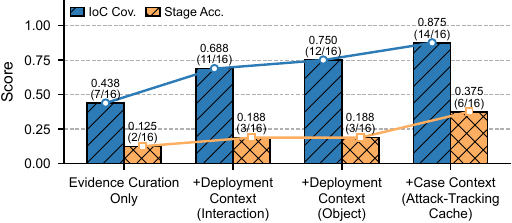}
		\caption{Dual-Context Investigation ablation on E3-CADETS with components added incrementally.}
		\label{fig:investigation_ablation}
	\end{figure}

	\noindent\textbf{Dual-Context Investigation.} Figure~\ref{fig:investigation_ablation} adds the dual-context components incrementally. With curated evidence alone, IoC coverage is 0.438 and stage-attribution accuracy is 0.125. Adding interaction baselines from the Deployment Context yields the largest single-step gain in IoC coverage, object baselines increase it further, and enabling the Case Context maintained by the Attack-Tracking Cache reaches the full-system results, indicating that the confidence-gated Case Context provides validated case state unavailable to independent fragment analysis.
	
	\noindent Table~\ref{tab:entity_evidence_trajectories} further traces representative entities through the dual-context investigation. Routine deployment activities may initially be labeled \texttt{suspicious}, but object confirmation downgrades them to \texttt{likely\_benign}, preventing their admission to the Attack-Tracking Cache and final narrative. In contrast, genuine attack entities are confirmed as \texttt{highly\_suspicious}, admitted with their stage assignments, and retained in the final narrative. These trajectories show that the Deployment Context filters deployment-specific benign activity, while the confidence-gated Attack-Tracking Cache prevents low-confidence judgments from propagating into subsequent investigation.
	
	\begin{table*}[t]
		\centering
		\caption{Entity-level evidence trajectories through the investigation pipeline. Routine deployment activities are contrasted with ground-truth attack IoCs.}
		\label{tab:entity_evidence_trajectories}
		\smallskip
		\fontsize{8}{10}\selectfont
		\setlength{\tabcolsep}{1.5pt}
		\begin{tabularx}{0.97\textwidth}{|>{\raggedright\arraybackslash}X|>{\centering\arraybackslash}p{0.055\textwidth}|>{\centering\arraybackslash}p{0.05\textwidth}|>{\centering\arraybackslash}p{0.065\textwidth}|>{\centering\arraybackslash}p{0.115\textwidth}|>{\centering\arraybackslash}p{0.12\textwidth}|>{\centering\arraybackslash}p{0.10\textwidth}|>{\centering\arraybackslash}p{0.07\textwidth}|>{\centering\arraybackslash}p{0.045\textwidth}|>{\centering\arraybackslash}p{0.12\textwidth}|}
			\hline
			\multirow{2}{*}{\textbf{Investigation stage}} & \multicolumn{5}{c|}{\textbf{Routine deployment activities}} & \multicolumn{4}{c|}{\textbf{Ground-truth attack IoCs}} \\
			\cline{2-10}
			& \texttt{vmstat} & \texttt{lsof} & \texttt{devctl} & \texttt{410.pkg-audit} & \texttt{411.pkg-backup} & \texttt{128.55.12.1} & \texttt{minions} & \texttt{XIM} & \texttt{53.158.101.118} \\
			\hline
			Fragment generation & \ding{51} & \ding{51} & \ding{51} & \ding{51} & \ding{51} & \ding{51} & \ding{51} & \ding{51} & \ding{51} \\
			\hline
			First-round LLM & {\texttt{S}} & {\texttt{LB}} & {\texttt{S}} & {\texttt{S}} & {\texttt{S}} & {\texttt{S}} & {\texttt{HS}} & {\texttt{HS}} & {\texttt{S}} \\
			\hline
			Object confirmation & {\texttt{LB}} & \texttt{N/A} & {\texttt{LB}} & {\texttt{LB}} & {\texttt{LB}} & {\texttt{HS}} & \texttt{N/A} & \texttt{N/A} & {\texttt{HS}} \\
			\hline
			Attack-Tracking Cache & \ding{55} & \ding{55} & \ding{55} & \ding{55} & \ding{55} & \ding{51}~\texttt{IR/C2} & \ding{51}~\texttt{IC/C2} & \ding{51}~\texttt{C2} & \ding{51}~\texttt{C2} \\
			\hline
			Final narrative & \ding{55} & \ding{55} & \ding{55} & \ding{55} & \ding{55} & \ding{51} & \ding{51} & \ding{51} & \ding{51} \\
			\hline
		\end{tabularx}
		\vspace{2pt}
		\parbox{\textwidth}{\scriptsize \ding{51}/\ding{55} denotes presence or absence at fragment generation and in the final narrative. Object confirmation reports the post-confirmation label (\S\ref{subsec:online-pipeline}); \texttt{N/A} indicates that confirmation was not triggered. The Attack-Tracking Cache row reports \ding{55} when no fragment containing the entity is admitted; otherwise, it reports \ding{51} followed by the accumulated stage assignment carried by the admitted fragment (\S\ref{subsec:reconstruction}).}
	\end{table*}
	
	\subsection{RQ6: Hyperparameter Sensitivity}
	\label{subsec:rq6}
	
	\noindent We answer RQ6 on E3-CADETS by varying one hyperparameter at a time while keeping the remaining settings at their defaults, covering the learning rate, Word2Vec dimension, GNN output dimension, TGN state dimension, and TGN neighborhood size. Across all tested configurations, recall remains $1.000$, so as a candidate generator \sysname retains all four labeled attack windows under every setting, and the variation appears only in precision. Precision degrades only under extreme learning rates or undersized GNN output dimensions, TGN state dimensions, and neighborhood sizes, which introduce additional false-positive windows, while all remaining settings keep precision at $1.000$ and the Word2Vec dimension shows no observed effect over the tested range. Figure~\ref{fig:hyperparameter_sensitivity} (Appendix~\ref{sec:appendix_hyper}) reports the complete per-parameter results.
	
\section{Related Work}
	\label{sec:related}
	
	We organize related work around three overlapping themes, namely provenance-based intrusion detection, attack investigation and reconstruction, and LLM-assisted security investigation.
	
\noindent\textbf{Provenance-based intrusion detection.}
A large body of work detects attacks by learning normal behavior from whole-system provenance graphs. Coarse-grained detectors such as StreamSpot~\cite{streamspot}, Unicorn~\cite{unicorn}, and ProGrapher~\cite{prographer} assign anomaly scores to execution graphs or temporal snapshots, while ThreaTrace~\cite{threatrace}, Flash~\cite{flash}, and MAGIC~\cite{magic} localize anomalous entities through node-level prediction or masked graph reconstruction. Orthrus~\cite{orthrus} combines temporal graph learning with causality analysis, ProvFusion~\cite{provfusion} fuses attribute-, structure-, and causality-based anomaly views, SLOT~\cite{qiao2025slot} applies graph reinforcement learning to localize APT activities, Kairos~\cite{kairos} predicts temporal edge types and combines reconstruction errors with node rarity, and VELOX~\cite{bilot2025sometimes} shows that a single-layer linear encoder can rival substantially more complex PIDSs. Regardless of granularity, these detectors emit fragmented local anomalies rather than evidence organized for investigation. \sysname likewise uses temporal edge reconstruction, but calibrates losses independently per relation and extends node rarity to \emph{(entity, relation, role)} keys, so that anomalies can be linked into continuous cross-window evidence queues (\S\ref{subsec:rq3}, \S\ref{subsec:rq5}).
	
\noindent\textbf{Attack investigation and reconstruction.}
A parallel line reconstructs the attack story rather than merely flagging elements. SLEUTH~\cite{sleuth} and HOLMES~\cite{holmes} correlate suspicious information flows into scenario graphs, NoDoze~\cite{nodoze} and PrioTracker~\cite{priotracker} prioritize alerts by event rarity, DepImpact~\cite{depimpact} back-propagates dependency impact to rank attack-relevant edges, and ATLAS~\cite{atlas} learns attack sequences from audit logs. Reconstruction also appears in recent detectors, with NodLink~\cite{nodlink} formalizing online investigation as a Steiner-tree problem, Kairos~\cite{kairos} summarizing anomalous windows into attack graphs, and Orthrus~\cite{orthrus} attributing detections to compact attack-relevant subgraphs. Across both lines, the output remains graphs or ranked event lists that a human analyst must still interpret, and each window or alert is reconstructed in isolation without accumulated case state. \sysname borrows the rarity insight of NoDoze and PrioTracker but elevates it to the relation-role level, and delivers a stage-structured natural-language narrative that carries validated attack state across windows.
	
\noindent\textbf{LLM-assisted security investigation.}
Most relevant to \sysname is OCR-APT~\cite{ocrapt}, the first system to use an LLM to generate human-readable APT reports from provenance subgraphs; because those subgraphs inherit upstream detector noise, it must add multi-stage validation and explicit hallucination filtering. Yet it neither grounds the LLM in the deployment's normal behavior nor maintains validated attack state across windows, the two missing contexts identified in \S\ref{sec:background}. \sysname supplies both, grounding judgment in the Deployment Context through retrieval-augmented generation~\cite{rag} and carrying the Case Context across windows, and achieves better measured investigation results under the same reported LLM backbone (\S\ref{subsec:rq1}).
	
	\section{Conclusion}
	\label{sec:conclusion}
	
	This work identified the missing contexts behind unreliable LLM-based attack investigation and presented \sysname, which curates anomalies into cross-window evidence queues, grounds LLM judgment in the Deployment Context, and keeps the investigation stateful through the confidence-gated Case Context, offering an effective design for reliable LLM investigation over system provenance. On six DARPA TC E3/E5 datasets, \sysname improves overall IoC recovery and attack-stage attribution over state-of-the-art provenance-based baselines, with the advantage persisting under a fixed LLM backbone and full-day investigations at dollar-level API cost.
	
	\bibliographystyle{plain}
	\bibliography{references}

@inproceedings{kairos,
  author    = {Zijun Cheng and Qiujian Lv and Jinyuan Liang and Yan Wang and Degang Sun and Thomas Pasquier and Xueyuan Han},
  title     = {Kairos: Practical Intrusion Detection and Investigation using Whole-system Provenance},
  booktitle = {Proceedings of the IEEE Symposium on Security and Privacy (S\&P)},
  year      = {2024},
}

@inproceedings{orthrus,
  author    = {Baoxiang Jiang and Tristan Bilot and Nour El Madhoun and Khaldoun Al Agha and Anis Zouaoui and Shahrear Iqbal and Xueyuan Han and Thomas Pasquier},
  title     = {{ORTHRUS}: Achieving High Quality of Attribution in Provenance-based Intrusion Detection Systems},
  booktitle = {Proceedings of the USENIX Security Symposium (USENIX Security)},
  year      = {2025},
}

@inproceedings{flash,
  author    = {Mati Ur Rehman and Hadi Ahmadi and Wajih Ul Hassan},
  title     = {Flash: A Comprehensive Approach to Intrusion Detection via Provenance Graph Representation Learning},
  booktitle = {Proceedings of the IEEE Symposium on Security and Privacy (S\&P)},
  year      = {2024},
}

@article{mikolov2013word2vec,
  author    = {Tomas Mikolov and Kai Chen and Greg Corrado and Jeffrey Dean},
  title     = {Efficient Estimation of Word Representations in Vector Space},
  journal   = {arXiv preprint arXiv:1301.3781},
  year      = {2013},
}

@inproceedings{tgn,
  author    = {Emanuele Rossi and Ben Chamberlain and Fabrizio Frasca and Davide Eynard and Federico Monti and Michael Bronstein},
  title     = {Temporal Graph Networks for Deep Learning on Dynamic Graphs},
  booktitle = {ICML Workshop on Graph Representation Learning and Beyond (GRL+)},
  year      = {2020},
}

@article{rgatconv,
  author    = {Dan Busbridge and Dane Sherburn and Pietro Cavallo and Nils Y. Hammerla},
  title     = {Relational Graph Attention Networks},
  journal   = {arXiv preprint arXiv:1904.05811},
  year      = {2019},
}

@inproceedings{nodoze,
  author    = {Wajih Ul Hassan and Shengjian Guo and Ding Li and Zhengzhang Chen and Kangkook Jee and Zhichun Li and Adam Bates},
  title     = {{NoDoze}: Combatting Threat Alert Fatigue with Automated Provenance Triage},
  booktitle = {Proceedings of the Network and Distributed System Security Symposium (NDSS)},
  year      = {2019},
}

@inproceedings{priotracker,
  author    = {Yushan Liu and Mu Zhang and Ding Li and Kangkook Jee and Zhichun Li and Zhenyu Wu and Junghwan Rhee and Prateek Mittal},
  title     = {Towards a Timely Causality Analysis for Enterprise Security},
  booktitle = {Proceedings of the Network and Distributed System Security Symposium (NDSS)},
  year      = {2018},
}

@inproceedings{streamspot,
  author    = {Emaad Manzoor and Sadegh M. Milajerdi and Leman Akoglu},
  title     = {Fast Memory-efficient Anomaly Detection in Streaming Heterogeneous Graphs},
  booktitle = {Proceedings of the ACM SIGKDD International Conference on Knowledge Discovery and Data Mining (KDD)},
  year      = {2016},
}

@inproceedings{unicorn,
  author    = {Xueyuan Han and Thomas Pasquier and Adam Bates and James Mickens and Margo Seltzer},
  title     = {{UNICORN}: Runtime Provenance-Based Detector for Advanced Persistent Threats},
  booktitle = {Proceedings of the Network and Distributed System Security Symposium (NDSS)},
  year      = {2020},
}

@article{threatrace,
  author    = {Su Wang and Zhiliang Wang and Tao Zhou and Hongbin Sun and Xia Yin and Dongqi Han and Han Zhang and Xingang Shi and Jiahai Yang},
  title     = {{THREATRACE}: Detecting and Tracing Host-Based Threats in Node Level Through Provenance Graph Learning},
  journal   = {IEEE Transactions on Information Forensics and Security},
  volume    = {17},
  pages     = {3972--3987},
  year      = {2022},
}

@inproceedings{magic,
  author    = {Zian Jia and Yun Xiong and Yuhong Nan and Yao Zhang and Jinjing Zhao and Mi Wen},
  title     = {{MAGIC}: Detecting Advanced Persistent Threats via Masked Graph Representation Learning},
  booktitle = {Proceedings of the USENIX Security Symposium (USENIX Security)},
  year      = {2024},
}

@inproceedings{provfusion,
  author    = {Fan Yang and Binyan Xu and Di Tang and Kehuan Zhang},
  title     = {Beyond Nodes vs. Edges: A Multi-View Fusion Framework for Provenance-Based Intrusion Detection},
  booktitle = {Proceedings of the IEEE Symposium on Security and Privacy (S\&P)},
  year      = {2026},
}

@inproceedings{ocrapt,
  author    = {Ahmed Aly and Essam Mansour and Amr M. Youssef},
  title     = {{OCR-APT}: Reconstructing {APT} Stories from Audit Logs using Subgraph Anomaly Detection and {LLMs}},
  booktitle = {Proceedings of the ACM SIGSAC Conference on Computer and Communications Security (CCS)},
  year      = {2025},
}

@inproceedings{holmes,
  author    = {Sadegh M. Milajerdi and Rigel Gjomemo and Birhanu Eshete and R. Sekar and V. N. Venkatakrishnan},
  title     = {{HOLMES}: Real-Time {APT} Detection through Correlation of Suspicious Information Flows},
  booktitle = {Proceedings of the IEEE Symposium on Security and Privacy (S\&P)},
  year      = {2019},
}

@incollection{hutchins2011intelligence,
  title     = {Intelligence-driven computer network defense informed by analysis of adversary campaigns and intrusion kill chains},
  author    = {Hutchins, Eric M. and Cloppert, Michael J. and Amin, Rohan M.},
  booktitle = {Leading Issues in Information Warfare \& Security Research},
  volume    = {1},
  pages     = {80--106},
  publisher = {Academic Publishing International},
  year      = {2011},
}

@inproceedings{sleuth,
  author    = {Md Nahid Hossain and Sadegh M. Milajerdi and Junao Wang and Birhanu Eshete and Rigel Gjomemo and R. Sekar and Scott Stoller and V. N. Venkatakrishnan},
  title     = {{SLEUTH}: Real-time Attack Scenario Reconstruction from {COTS} Audit Data},
  booktitle = {Proceedings of the USENIX Security Symposium (USENIX Security)},
  year      = {2017},
}

@inproceedings{atlas,
  author    = {Abdulellah Alsaheel and Yuhong Nan and Shiqing Ma and Le Yu and Gregory Walkup and Z. Berkay Celik and Xiangyu Zhang and Dongyan Xu},
  title     = {{ATLAS}: A Sequence-based Learning Approach for Attack Investigation},
  booktitle = {Proceedings of the USENIX Security Symposium (USENIX Security)},
  year      = {2021},
}

@inproceedings{depimpact,
  author    = {Pengcheng Fang and Peng Gao and Changlin Liu and Erman Ayday and Kangkook Jee and Ting Wang and Yanfang Ye and Zhuotao Liu and Xusheng Xiao},
  title     = {Back-Propagating System Dependency Impact for Attack Investigation},
  booktitle = {Proceedings of the USENIX Security Symposium (USENIX Security)},
  year      = {2022},
}

@inproceedings{nodlink,
  author    = {Shaofei Li and Feng Dong and Xusheng Xiao and Haoyu Wang and Fei Shao and Jiedong Chen and Yao Guo and Xiangqun Chen and Ding Li},
  title     = {{NODLINK}: An Online System for Fine-Grained {APT} Attack Detection and Investigation},
  booktitle = {Proceedings of the Network and Distributed System Security Symposium (NDSS)},
  year      = {2024},
}

@inproceedings{sigl,
  author    = {Xueyuan Han and Xiao Yu and Thomas Pasquier and Ding Li and Junghwan Rhee and James Mickens and Margo Seltzer and Haifeng Chen},
  title     = {{SIGL}: Securing Software Installations Through Deep Graph Learning},
  booktitle = {Proceedings of the USENIX Security Symposium (USENIX Security)},
  year      = {2021},
}

@inproceedings{prographer,
  author    = {Fan Yang and Jiacen Xu and Chunlin Xiong and Zhou Li and Kehuan Zhang},
  title     = {{PROGRAPHER}: An Anomaly Detection System based on Provenance Graph Embedding},
  booktitle = {Proceedings of the USENIX Security Symposium (USENIX Security)},
  year      = {2023},
}

@inproceedings{rag,
  author    = {Patrick Lewis and Ethan Perez and Aleksandra Piktus and Fabio Petroni and Vladimir Karpukhin and Naman Goyal and Heinrich K{\"u}ttler and Mike Lewis and {Wen-tau} Yih and Tim Rockt{\"a}schel and Sebastian Riedel and Douwe Kiela},
  title     = {Retrieval-Augmented Generation for Knowledge-Intensive {NLP} Tasks},
  booktitle = {Advances in Neural Information Processing Systems (NeurIPS)},
  year      = {2020},
}

@inproceedings{bilot2025sometimes,
  title={Sometimes Simpler is Better: A Comprehensive Analysis of {State-of-the-Art} {Provenance-Based} Intrusion Detection Systems},
  author={Bilot, Tristan and Jiang, Baoxiang and Li, Zefeng and El Madhoun, Nour and Al Agha, Khaldoun and Zouaoui, Anis and Pasquier, Thomas},
  booktitle={Proceedings of the USENIX Security Symposium (USENIX Security)},
  year={2025}
}

@misc{etw,
  title={Event Tracing for Windows ({ETW})},
  author={{Microsoft}},
  howpublished={\url{https://learn.microsoft.com/en-us/windows-hardware/drivers/devtest/event-tracing-for-windows--etw-}},
  note={Accessed 29th January 2025},
  year={2021}
}

@misc{linuxaudit,
  title={{auditd(8)} - {Linux} Manual Page},
  author={Grubb, Steve},
  howpublished={\url{https://man7.org/linux/man-pages/man8/auditd.8.html}},
  note={Accessed 29th January 2025},
  year={2021}
}

@inproceedings{pasquier2017practical,
  title={Practical Whole-System Provenance Capture},
  author={Pasquier, Thomas and Han, Xueyuan and Goldstein, Mark and Moyer, Thomas and Eyers, David and Seltzer, Margo and Bacon, Jean},
  booktitle={Proceedings of the ACM Symposium on Cloud Computing (SoCC)},
  year={2017}
}

@techreport{mitre_attack,
  author    = {Blake E. Strom and Andy Applebaum and Doug P. Miller and Kathryn C. Nickels and Adam G. Pennington and Cody B. Thomas},
  title     = {{MITRE} {ATT\&CK}: Design and Philosophy},
  institution = {The MITRE Corporation},
  year      = {2020},
  note      = {\url{https://attack.mitre.org/}}
}

@inproceedings{dong2023we,
  title={Are we there yet? An industrial viewpoint on provenance-based endpoint detection and response tools},
  author={Dong, Feng and Li, Shaofei and Jiang, Peng and Li, Ding and Wang, Haoyu and Huang, Liangyi and Xiao, Xusheng and Chen, Jiedong and Luo, Xiapu and Guo, Yao and Chen, Xiangqun},
  booktitle={Proceedings of the ACM SIGSAC Conference on Computer and Communications Security (CCS)},
  year={2023}
}

@inproceedings{goyal2024r,
  title={{R-CAID}: Embedding Root Cause Analysis within Provenance-Based Intrusion Detection},
  author={Akul Goyal and Gang Wang and Adam Bates},
  booktitle={Proceedings of the IEEE Symposium on Security and Privacy (S\&P)},
  year={2024}
}

@inproceedings{king2023edgetorrent,
  title={{EdgeTorrent}: Real-time Temporal Graph Representations for Intrusion Detection},
  author={Isaiah J. King and Xiaokui Shu and Jiyong Jang and Kevin Eykholt and Taesung Lee and H. Howie Huang},
  booktitle={Proceedings of the International Symposium on Research in Attacks, Intrusions and Defenses (RAID)},
  year={2023}
}

@article{xu2026deepseek,
  title={{DeepSeek-V4}: Towards Highly Efficient Million-Token Context Intelligence},
  author={{DeepSeek-AI} and Anyi Xu and Bangcai Lin and Bing Xue and Bingxuan Wang and Bingzheng Xu and Bochao Wu and Bowei Zhang and Chaofan Lin and Chen Dong and Chenchen Ling and others},
  journal={arXiv preprint arXiv:2606.19348},
  year={2026}
}

@inproceedings{kohno2023ethical,
  title={Ethical frameworks and computer security trolley problems: Foundations for conversations},
  author={Kohno, Tadayoshi and Acar, Yasemin and Loh, Wulf},
  booktitle={Proceedings of the USENIX Security Symposium (USENIX Security)},
  year={2023}
}

@misc{qdrant,
  title={Qdrant Vector Database},
  author={{Qdrant}},
  howpublished={\url{https://qdrant.tech/documentation/}},
  note={Accessed July 2025},
  year={2025}
}

@misc{darpa_e3,
  title={Transparent Computing Engagement 3 Data Release},
  author={{DARPA I2O}},
  howpublished={\url{https://github.com/darpa-i2o/Transparent-Computing/blob/master/README-E3.md}},
  note={Accessed 29th January 2025},
  year={2018}
}

@misc{darpa_e5,
  title={Transparent Computing Engagement 5 Data Release},
  author={{DARPA I2O}},
  howpublished={\url{https://github.com/darpa-i2o/Transparent-Computing/blob/master/README.md}},
  note={Engagement conducted in 2019; data release published in 2020. Accessed 29th January 2025},
  year={2019}
}

@misc{tc_gt_e3,
  title={{TA5.1 Ground Truth Report Engagement 3}},
  author={{Kudu Dynamics}},
  howpublished={DARPA Transparent Computing},
  month=may,
  year={2018},
  note={Distributed with the Engagement 3 data release: \url{https://github.com/darpa-i2o/Transparent-Computing/blob/master/README-E3.md}}
}

@misc{ta51_final_e5,
  title={{TA5.1 Final Report Engagement 5}},
  author={{Kudu Dynamics}},
  howpublished={DARPA Transparent Computing},
  year={2019},
  note={Revision 1.0, June 28, 2019. Distributed with the Engagement 5 data release: \url{https://github.com/darpa-i2o/Transparent-Computing/blob/master/README.md}}
}

@inproceedings{qiao2025slot,
  title={Slot: Provenance-Driven {APT} detection through graph reinforcement learning},
  author={Qiao, Wei and Feng, Yebo and Li, Teng and Ma, Zhuo and Shen, Yulong and Ma, Jianfeng and Liu, Yang},
  booktitle={Proceedings of the ACM SIGSAC Conference on Computer and Communications Security (CCS)},
  year={2025}
}

@inproceedings{rgcn,
  title={Modeling relational data with graph convolutional networks},
  author={Schlichtkrull, Michael and Kipf, Thomas N. and Bloem, Peter and van den Berg, Rianne and Titov, Ivan and Welling, Max},
  booktitle={Proceedings of the European Semantic Web Conference (ESWC)},
  year={2018},
  organization={Springer}
}

@article{unimp,
  title={Masked label prediction: Unified message passing model for semi-supervised classification},
  author={Shi, Yunsheng and Huang, Zhengjie and Feng, Shikun and Zhong, Hui and Wang, Wenjing and Sun, Yu},
  journal={arXiv preprint arXiv:2009.03509},
  year={2020}
}

@article{gat,
  title={Graph attention networks},
  author={Veli{\v{c}}kovi{\'c}, Petar and Cucurull, Guillem and Casanova, Arantxa and Romero, Adriana and Li{\`o}, Pietro and Bengio, Yoshua},
  journal={arXiv preprint arXiv:1710.10903},
  year={2017}
}

@article{liu2023prompt,
  title={Prompt injection attack against {LLM}-integrated applications},
  author={Liu, Yi and Deng, Gelei and Li, Yuekang and Wang, Kailong and Wang, Zihao and Wang, Xiaofeng and Zhang, Tianwei and Liu, Yepang and Wang, Haoyu and Zheng, Yan and others},
  journal={arXiv preprint arXiv:2306.05499},
  year={2023}
}

@article{shi2023badgpt,
  title={{BadGPT}: Exploring security vulnerabilities of {ChatGPT} via backdoor attacks to {InstructGPT}},
  author={Shi, Jiawen and Liu, Yixin and Zhou, Pan and Sun, Lichao},
  journal={arXiv preprint arXiv:2304.12298},
  year={2023}
}

@article{zou2023universal,
  title={Universal and transferable adversarial attacks on aligned language models},
  author={Zou, Andy and Wang, Zifan and Carlini, Nicholas and Nasr, Milad and Kolter, J. Zico and Fredrikson, Matt},
  journal={arXiv preprint arXiv:2307.15043},
  year={2023}
}

@article{hubinger2024sleeper,
  title={Sleeper agents: Training deceptive {LLM}s that persist through safety training},
  author={Hubinger, Evan and Denison, Carson and Mu, Jesse and Lambert, Mike and Tong, Meg and MacDiarmid, Monte and Lanham, Tamera and Ziegler, Daniel M and Maxwell, Tim and Cheng, Newton and others},
  journal={arXiv preprint arXiv:2401.05566},
  year={2024}
}

@inproceedings{zou2025poisonedrag,
  title={{PoisonedRAG}: Knowledge corruption attacks to retrieval-augmented generation of large language models},
  author={Zou, Wei and Geng, Runpeng and Wang, Binghui and Jia, Jinyuan},
  booktitle={Proceedings of the USENIX Security Symposium (USENIX Security)},
  year={2025}
}
	
	\appendix
	\section{Ethics Considerations}
	\label{sec:appendix_ethics}
	In conducting this research, we have carefully considered its ethical implications~\cite{kohno2023ethical}. To the best of our knowledge, this work raises no ethical issues. All experiments are conducted solely on publicly released benchmark datasets that were collected in an ethical manner and contain no sensitive information.
	
	\section{Compliance with the Open Science Policy}
	\label{sec:appendix_openscience}
	We open-source the complete implementation of \sysname at \url{https://anonymous.4open.science/r/Anchor-3619/}, together with all evaluation labels used in this study, including the window-level labels of the six datasets (with our corrections to the public Kairos annotations and their justifications), the IoC labels, and the stage-attribution labels, as well as the configurations and scripts required to reproduce our experiments. All of these artifacts will be made publicly available upon paper acceptance.
	
	\section{Ground Truth Label Corrections}
	\label{sec:appendix_labels}
	
	Table~\ref{tab:label_corrections} summarizes our corrections and supplements to the ground truth labels used in window-level evaluation. All modifications are strictly based on the official engagement reports (TA5.1 Final Report for E5 datasets, TA5.1 Ground Truth Report for E3 datasets). We mark each modified window with its action type. \emph{Added} denotes a previously unlabeled attack window that we include based on report evidence, \emph{Deleted} denotes a window originally labeled as malicious but lacking any corresponding attack activity in the report, and \emph{Corrected} denotes a window whose temporal boundaries are adjusted to align with the actual attack timeline.
	
	\begin{table*}[t]
		\centering
		\caption{Ground truth label corrections and supplements. \emph{Added} marks an attack window missing from original labels, \emph{Deleted} marks an originally labeled window with no corresponding attack activity, and \emph{Corrected} marks a window with adjusted temporal boundaries.}
		\label{tab:label_corrections}
		\smallskip
		\small
		\begin{tabular}{@{}ccclp{7.5cm}@{}}
			\toprule
			Dataset & Date & Time Range & Action & Attack Activity \\
			\midrule
			\multirow{5}{*}{E5-CADETS}
			& 5/17 & 10:48--11:03 & Added & Exploit success, C2 session established, recon commands \\
			& 5/17 & 11:03--11:19 & Added & Dual C2 sessions maintained (cadets-1 + cadets-2) \\
			& 5/17 & 11:19--11:34 & Added & Data exfiltration: \texttt{cat /etc/passwd} on cadets-1 \\
			& 5/17 & 14:06--14:22 & Added & Missed C2 listener discovered, connections closed \\
			& 5/17 & 15:22--15:38 & Added & Final C2 status check, all connections terminated \\
			\cmidrule{1-5}
			\multirow{8}{*}{E5-ClearScope}
			& 5/15 & 14:40--14:55 & Added & Barephone APK malicious activity final phase (ends 14:42) \\
			& 5/15 & 15:25--15:40 & Added & Appstarter APK installation (15:39), attack chain start \\
			& 5/17 & 11:48--12:03 & Added & Firefox Drakon exploit and payload execution \\
			& 5/17 & 14:18--14:33 & Added & MyApp APK first installation attempt (14:27) \\
			& 5/17 & 14:33--14:48 & Added & MyApp multiple installation retries \\
			& 5/17 & 15:18--15:34 & Added & MyApp persistent retry activity \\
			& 5/17 & 15:49--16:04 & Added & Lockwatch APK malicious activity (Java APT success) \\
			& 5/17 & 16:04--16:19 & Added & MyApp retry activity continues \\
			\cmidrule{1-5}
			E3-THEIA
			& 4/10 & 13:46--14:02 & Added & Exploit retry period (Firefox crash, kill, retry with new target) \\
			\cmidrule{1-5}
			\multirow{3}{*}{E5-THEIA}
			& 5/15 & 13:58--14:13 & Deleted & No attack activity (50 min before earliest reported attack) \\
			& 5/15 & 14:43--14:58 & Corrected & Firefox exploit, C2 connection, privilege escalation to root \\
			& 5/15 & 14:58--15:15 & Corrected & Process injection into sshd, persistence established \\
			\bottomrule
		\end{tabular}
	\end{table*}
	
\section{Provenance Graph Schema}
\label{sec:appendix_schema}

Table~\ref{tab:graph} summarizes the provenance graph schema used by \sysname, covering the modeled entity types, their interaction relations, and the entity attributes retained for feature extraction.

\begin{table}[t]
	\centering
	\caption{Provenance graph schema.}
	\label{tab:graph}
	\footnotesize
	\setlength{\tabcolsep}{4pt}
	\begin{tabularx}{\columnwidth}{|l|X|l|}
		\hline
		\textbf{Events} & \textbf{Relations} & \textbf{Entity Attributes} \\
		\hline
		Process $\leftrightarrow$ Process & Start, Close, Clone & name, cmd \\
		\hline
		Process $\leftrightarrow$ File & Read, Write, Open, Execute & path \\
		\hline
		Process $\leftrightarrow$ NetFlow & Send, Receive & src/dst IP, port \\
		\hline
	\end{tabularx}
\end{table}

\section{Dataset Splitting}
\label{sec:appendix_split}
	
	Table~\ref{tab:data_split} summarizes the data splitting for each dataset. Training days are used to train the detector and construct the deployment normalcy baselines. Validation days are used to calibrate detection thresholds and normalcy statistics. Evaluation days contain the attack activity against which we measure detection and investigation performance.
	
	\begin{table}[t]
		\centering
		\caption{Dataset splitting details for each dataset.}
		\label{tab:data_split}
		\smallskip
		\small
		
		\resizebox{\columnwidth}{!}{%
			\begin{tabular}{|l|l|l|l|}
				\hline
				Dataset & Training Days & Validation Days & Evaluation Days \\
				\hline
				E3-CADETS & 2018-04-02/03/04 & 2018-04-05 & 2018-04-06/07/12/13 \\
				\hline
				E5-CADETS & 2019-05-08/09/11 & 2019-05-12 & 2019-05-15/16/17 \\
				\hline
				E3-THEIA & 2018-04-03/04/05 & 2018-04-09 & 2018-04-10/11/12 \\
				\hline
				E5-THEIA & 2019-05-08/09 & 2019-05-11 & 2019-05-14/15 \\
				\hline
				E3-ClearScope & 2018-04-04/05/06 & 2018-04-07 & 2018-04-10/11 \\
				\hline
				E5-ClearScope & 2019-05-08/09/11 & 2019-05-12 & 2019-05-14/15/17 \\
				\hline
			\end{tabular}%
		}
		
	\end{table}
	
\section{Hyperparameter Sensitivity Results}
\label{sec:appendix_hyper}

\begin{figure*}[t]
	\centering
	\includegraphics[width=0.55\textwidth]{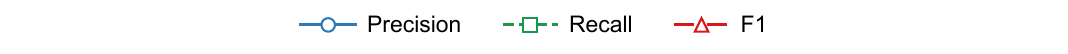}\par\vspace{2pt}
	\begin{subfigure}[t]{0.19\textwidth}
		\centering
		\includegraphics[width=\linewidth]{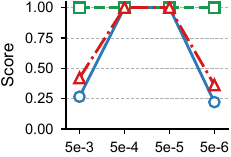}
		\caption{Learning rate}
		\label{fig:hyper_lr}
	\end{subfigure}\hfill
	\begin{subfigure}[t]{0.19\textwidth}
		\centering
		\includegraphics[width=\linewidth]{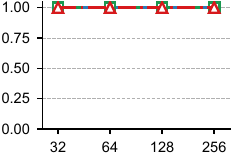}
		\caption{Word2Vec dim.}
		\label{fig:hyper_w2v}
	\end{subfigure}\hfill
	\begin{subfigure}[t]{0.19\textwidth}
		\centering
		\includegraphics[width=\linewidth]{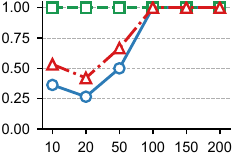}
		\caption{GNN output dim.}
		\label{fig:hyper_edge}
	\end{subfigure}\hfill
	\begin{subfigure}[t]{0.19\textwidth}
		\centering
		\includegraphics[width=\linewidth]{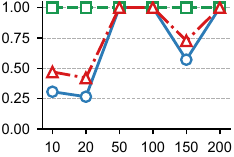}
		\caption{TGN state dim.}
		\label{fig:hyper_state}
	\end{subfigure}\hfill
	\begin{subfigure}[t]{0.19\textwidth}
		\centering
		\includegraphics[width=\linewidth]{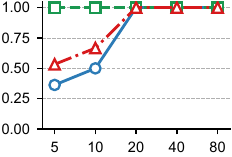}
		\caption{TGN neighborhood size}
		\label{fig:hyper_neighbor}
	\end{subfigure}
	\caption{Hyperparameter sensitivity on E3-CADETS. Each subplot varies one parameter and reports precision, recall, and F1.}
	\label{fig:hyperparameter_sensitivity}
\end{figure*}

Figure~\ref{fig:hyperparameter_sensitivity} reports the complete per-parameter results of the hyperparameter sensitivity study in \S\ref{subsec:rq6}, varying one hyperparameter at a time on E3-CADETS while keeping the remaining settings at their defaults.

\section{Case Study on E5-CADETS}
\label{sec:appendix_casestudy}
	
	{We use the Nginx Drakon APT attack in E5-CADETS as a case study of the attack narrative reconstructed by \sysname. The attack spans May 16 and 17, 2019 and targets two FreeBSD hosts (128.55.12.51 and 128.55.12.75). The attacker triggers an \texttt{nginx} backdoor through malicious HTTP requests to gain initial access, performs reconnaissance with commands such as \texttt{hostname}, \texttt{whoami}, and \texttt{cat /etc/passwd}, and steals \texttt{/etc/passwd}. The activity on May 17 reruns the previous day's attack, where the attacker first fails to deliver the payload via 98.23.182.25 and then establishes a command-and-control channel through 128.55.12.233~\cite{ta51_final_e5}.}
	
	{The evidence curation layer of \sysname submits 23 candidate windows (W1 to W23) to the grounded investigator. Evidence fragments in 14 of them are judged highly suspicious and enter the final attack narrative, with W1 to W6 on May 16 and W16 to W23 on May 17, while the remaining nine candidate windows are excluded after investigation reveals no attack evidence. Among the 15 ground-truth IoCs, \sysname recovers 13, of which 10 are attributed to correct kill-chain stages, corresponding to the coverage of 0.867 and stage-attribution accuracy of 0.667 reported in Table~\ref{tab:rq2_detection_evidence}.}
	
\begin{figure}[t]
		\centering
		\includegraphics[width=\columnwidth]{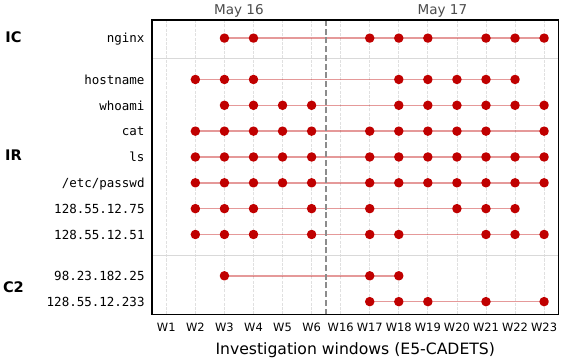}
		\caption{{Activity footprints of the 10 ground-truth IoCs correctly attributed by \sysname on E5-CADETS, grouped by their kill-chain stages. Dots mark the investigation windows in which each entity appears in the final report.}}
		\label{fig:casestudy_timeline}
	\end{figure}

	{Figure~\ref{fig:casestudy_timeline} shows the activity footprints of the 10 correctly attributed IoCs across the investigated windows. The attack pauses on the evening of May 16 and resumes the next morning, and the case state maintained by the Attack-Tracking Cache allows \sysname to organize the temporally dispersed evidence of both days into a single attack case. \texttt{nginx} is flagged and attributed as early as May 16 (W3 and W4). When the attack reruns on May 17 and the new C2 address 128.55.12.233 appears, its communication with \texttt{nginx} is directly attributed to the command-and-control stage of the existing case, and the final report presents this cross-day attack in a complete kill-chain structure.}
	
\section{LLM Prompt Templates}
\label{sec:appendix_prompts}

This appendix reproduces the two core system prompts used across all datasets. Each LLM request combines one fixed system prompt shown below with a JSON user message carrying runtime evidence. Two distinct placeholders preserve the two runtime concatenations used by the implementation.

\begin{promptbox}{Core Prompt 1. \texttt{FRAGMENT\_SYSTEM\_PROMPT}}
	\raggedright
	You are a cyber attack investigation analyst. Your task is to compare the current suspicious fragment against the retrieved normal baselines and the provided high-risk process-file rule list, and determine whether the fragment is suspicious.
	
	\medskip
	You will receive:
	
	1. A process-centered suspicious fragment.\\
	2. The fragment summary.\\
	3. Retrieved interaction baseline evidence.\\
	4. A high-risk process-file rule list.\\
	5. Optional prior high-confidence investigation memory from earlier windows in the same investigation queue.\\
	6. Optional object baseline evidence, added only when the first-round result is suspicious and further confirmation is needed.
	
	\medskip
	Your goals are:
	
	- determine whether the fragment is suspicious\\
	- identify which interactions are suspicious\\
	- identify which interactions are likely benign\\
	- determine which APT lifecycle stage the fragment most likely corresponds to
	
	\medskip
	You must follow these rules:
	
	- prioritize interaction baseline evidence as the primary grounding context\\
	- use the high-risk process-file rule list as semantic risk context instead of exact rule matching\\
	- use prior high-confidence investigation memory as continuity context, not as a replacement for current evidence; when investigation\_memory includes prior confirmed lifecycle stages, use the stage progression as auxiliary context to inform your judgment, but always prioritize baseline evidence\\
	- use object baseline evidence only as supplementary confirmation\\
	- treat time information as weak evidence only\\
	- do not mark an action as malicious only because the process name looks suspicious\\
	- do not mark an action as benign only because the process is common\\
	- base your judgment on deviation from normal interaction baselines and semantic consistency with the provided high-risk rule list\\
	- in reasoning\_summary, explicitly state which interaction baselines, which high-risk rules, which prior investigation memory items, and which optional object baselines were used\\
	- return JSON only
	
	\medskip
	\underline{\texttt{<APT\_LIFECYCLE\_STAGE\_DEFINITIONS>}}
	
	\medskip
	You must return JSON with exactly the following fields:
	
	- fragment\_label\\
	- suspicious\_interactions\\
	- benign\_interactions\\
	- reasoning\_summary\\
	- lifecycle\_stage
	
	\medskip
	Requirements:
	
	- fragment\_label must be one of: highly\_suspicious, suspicious, likely\_benign, uncertain\\
	- suspicious\_interactions must list the interactions judged as suspicious\\
	- benign\_interactions must list the interactions judged as likely benign\\
	- reasoning\_summary must explicitly mention which interaction baselines, which high-risk rules, which prior investigation memory items, and which optional object baselines were used\\
	- lifecycle\_stage must be one of: \underline{\texttt{<APT\_LIFECYCLE\_STAGES>}}
\end{promptbox}

\begin{promptbox}{Core Prompt 2. \texttt{STORY\_UPDATE\_SYSTEM\_PROMPT}}
	\raggedright
	You are a cyber attack investigation analyst. You will receive the previous global attack story, the previous confirmed attack stages, and the current window summary.
	
	\medskip
	Your task is to update the global attack story incrementally.
	
	\medskip
	\underline{\texttt{<APT\_LIFECYCLE\_STAGE\_DEFINITIONS>}}
	
	\medskip
	You must:
	
	- decide whether the current window continues the existing attack chain\\
	- decide whether the current window introduces a new attack stage (new\_stage); if no new stage, set new\_stage to "none"\\
	- update confirmed\_stages: take the union of previous\_confirmed\_stages and the current window's covered\_stages, keeping only stages with confirmed evidence\\
	- identify the most important entities introduced or reinforced by this window\\
	- update the global attack story\\
	- explicitly ground the update in the previous story and the current window summary\\
	- return JSON only
	
	\medskip
	You must return JSON with exactly the following fields:
	
	- attack\_continues\\
	- new\_stage\\
	- confirmed\_stages\\
	- key\_entities\\
	- updated\_story\\
	- reasoning\_summary
	
	\medskip
	Requirements:
	
	- new\_stage must be one of: \underline{\texttt{<APT\_LIFECYCLE\_STAGES>}}\\
	- confirmed\_stages must be a list of stages from the APT lifecycle stages list above
\end{promptbox}

\end{document}